\documentclass[mathpazo]{cicp}

\usepackage{booktabs,longtable}
\usepackage{graphicx}
\usepackage{dcolumn}
\usepackage{bm}
\usepackage{subfigure}
\usepackage{makecell, rotating}
\usepackage{epsfig}
\usepackage{wrapfig}
\usepackage{float}
\usepackage{algorithm}
\usepackage{algorithmic}
\usepackage{color}
\usepackage{bbm}

\begin{document}
\title[Stability of Soft Quasicrystals]{Stability of Soft Quasicrystals in a Coupled-Mode
Swift-Hohenberg Model for Three-Component Systems}



\author[Jiang et.~al.]{Kai Jiang\affil{1},
 Jiajun Tong\affil{2}, and Pingwen Zhang\affil{3}\comma\corrauth}
 \address{
   \affilnum{1}\ School of Mathematics
   and Computational Science, Xiangtan University, Hunan, P.R.
   China, 411105.
   \\
   \affilnum{2}\ School of Mathematical Sciences,
   Peking University, Beijing P.R. China, 100871.
   \\
   \affilnum{3}\ LMAM, CAPT and School of Mathematical Sciences,
   Peking University, Beijing P.R. China, 100871.
   }
 \emails{
 {\tt kaijiang@xtu.edu.cn} (K.~Jiang),
  {\tt jiajun.tong@nyu.edu} (J.~Tong),
 {\tt  pzhang@pku.edu.cn} (P.~Zhang)}


\begin{abstract}
In this article, we discuss the stability of soft quasicrystalline
phases in a coupled-mode Swift-Hohenberg model for
three-component systems, where the characteristic length scales
are governed by the positive-definite gradient terms.
Classic two-mode approximation method and
direct numerical minimization are applied to the model. In the
latter approach, we apply the projection method to deal with the
potentially quasiperiodic ground states. A variable cell
method of optimizing the shape and size of higher-dimensional
periodic cell is developed to minimize the free energy with
respect to the order parameters. Based on the developed numerical
methods, we rediscover decagonal and dodecagonal quasicrystalline
phases, and find diverse periodic phases and complex modulated phases.
Furthermore, phase diagrams are obtained in
various phase spaces by comparing the free energies of different
candidate structures.
It does show not only the important roles of
system parameters, but also the effect of optimizing
computational domain. In particular, the optimization of
computational cell allows us to capture the ground states and
phase behavior with higher fidelity. We also make some
discussions on our results and show the potential of applying our
numerical methods to a larger class of mean-field free energy functionals.
\end{abstract}

\pac{61.44.Br, 64.75.Yz, 64.70.Km, 82.70.-y}

\maketitle

\section{Introduction\label{Section: Introduction}}
Since the first discovery of quasicrystal in a rapidly-quenched Al-Mn alloy in 1980s \cite{shechtman1984metallic}, hundreds of metallic quasicrystals, whose building blocks are on the atomic scale, are found both in syntheses in the laboratory \cite{he1988decagonal,tsai1989stable,steurer2009crystallography,steurer2004twenty,fujiwara2007quasicrystals,tsai2008icosahedral} and in nature \cite{bindi2009natural}.
More recently, a growing number of mesoscopic quasicrystalline
orders are reported in soft-matter systems
\cite{zeng2004supramolecular,hayashida2007polymeric,dotera2011quasicrystals,
zhang2012dodecagonal}.
These soft quasicrystals exhibit distinct properties from their solid-state counterparts.
For example, dodecagonal or even higher order symmetry are
possible in the soft-matter systems \cite{zeng2004supramolecular,
hayashida2007polymeric, fischer2011colloidal}, while dodecagonal solid-state quasicrystals are rarely seen \cite{steurer2004twenty}.
Hence, soft quasicrystals are thought to have essentially different
source of stability.


Theoretical studies on the origin and stability of
an order pattern, including periodic and quasiperiodic crystals,
often involve minimizing a suitable free energy functional of
the system, and comparing free energies of different
candidate phases \cite{chaikin1995principles,
archer2013quasicrystalline}.
Therefore a systematic investigation of the stability of quasicrystals
requires the availability of appropriate free energy functionals
and accurate methods of evaluating free energies of quasicrystals.
Several microscopic models have been studied over the years to
explore the formation of quasicrystals arising from pair
potentials with more than one microscopic length scales. Among
them, Denton and L{\"o}wen
\cite{denton1998stability} obtained stable colloidal
quasicrystals within one-component system.
On the other hand, phenomenological models based on
coarse-grained free energy functionals
are widely applied and particularly useful to the study of the
stability of soft quasicrystals
\cite{alexander1978should,bak1985phenomenological,jaric1985long,kalugin1985_86,gronlund1988instability,mermin1985mean,dotera2007mean}.
The earliest one can trace back to Alexander and McTague \cite{alexander1978should}.
They showed the possibility of stabilizing icosahedral quasicrystal using a Landau-type free energy functional with one order parameter.
Related works include Bak \cite{bak1985phenomenological}, Jari{\'c} \cite{jaric1985long}, Kalugin \emph{et al.}~\cite{kalugin1985_86} and Gronlund and Mermin \cite{gronlund1988instability}.
Mermin and Troian \cite{mermin1985mean} followed Alexander and
McTague's theory, but introduced a second order parameter to obtain stable icosahedral quasicrystal.
Swift and Hohenberg \cite{swift1977hydrodynamic} added a
positive-definite gradient term into the free energy functional
to represent the effect of characteristic length scale.
From the viewpoint of Fourier space, the gradient term is small only near the
critical wave number $k_c=1$, thus suppressing the growth of any
instabilities with wave numbers away from this value.
Their model can be used to describe the supercritical instability transition from a
homogeneous state to one-mode patterns.
After that, M{\"u}ller \cite{muller1994model} used a set of two coupled
partial differential equations; the pattern of a primary
field is stabilized by a secondary coupled field which
provides an effective space-dependent forcing.
Two-dimensional quasicrystals, consisting of 8- and 12-fold
orientational order have been obtained in their models.
Later, Dotera \cite{dotera2007mean} extended Mermin-Troian model
to ABC star copolymer systems with incompressible condition;
several ordered two-dimensional phases were investigated,
including quasicrystals with decagonal and dodecagonal symmetry
and an Archimedean tiling pattern named $(3.3.4.3.4)$.

Beyond the study of soft quasicrystals, Lifshitz and Petrich
\cite{lifshitz1997theoretical} investigated quasicrystalline patterns arising in parametrically-excited surface waves.
Although motivated by different physical phenomena, the free energy functional of their model is in a similar form with those in the preceding Landau-type models.
They used only one order parameter describing the amplitude of the standing-wave pattern, yet an additional differential term characterizing multiple-frequency forcing in the free energy functional.
They successfully stabilized dodecagonal quasicrystalline pattern
by introducing multiple-frequency forcing and three-body interactions.

Although the forementioned theoretical studies focus on different physical issues,
they reveal the same physical nature in the formation of quasicrystals. (i) They
may arise from the competition of multiple length scales \cite{muller1994model,
lifshitz1997theoretical, denton1998stability, archer2013quasicrystalline, mermin1985mean, dotera2007mean}.
(ii) At the same time the three-body interactions characterized by the
cubic terms play an important role in stabilizing soft quasicrystals
\cite{lifshitz2007soft}. Therefore constructing an appropriate
free energy functional to study the stability of quasicrystals
should consist of more than one length scale and three-body
interactions at least.
It is also noticed that the soft quasicrystals are usually discovered in
relatively complicated multicomponent soft-matter systems,
such as in ABC star-shaped terpolymers
\cite{hayashida2007polymeric}, ABA'C tetrablock copolymers
\cite{zhang2012dodecagonal}. Therefore a free energy functional
with multiple coupled fields may be appropriate to study these
complex soft-matter systems.
In this work, we will make full use of those underlying physical
nature of quasicrystals in the literature and extend the
Swift-Hohenberg model to study the stability of soft quasicrystals.
The extension is twofold: one is
a positive-definite gradient terms with multi-mode
interaction to describe competing length scales; the other one is
that we introduce multiple order parameters which would be available to study the generic
multicomponent systems.

Besides a proper free energy functional,
the study of thermodynamic stability requires accurate
evaluation of free energy of various ordered phases.
A number of methods have been developed to calculate energy of
quasicrystalline pattens, including two-mode approximation approach
\cite{chaikin1995principles, lifshitz1997theoretical},
using periodic structures to approximate
quasicrystals with large unit cells
\cite{lifshitz1997theoretical, reinhardt2013computing,
engel2007self, dotera2007mean, muller1994model}.
An alternative approach to calculate the free energy of
quasicrystals is based on the observation that quasiperiodic
lattices can be generated by a cut-and-project method from
higher-dimensional periodic lattices\,\cite{janot1992quasicrystals,
steinhardt1991quasicrysrtals, goldman1993quasicrystals}.
It follows that the density and free energy of quasicrystals can be obtained using
the quasiperiodic lattices derived from the higher-dimensional
periodic structure. A approach along this line is the
Gaussian method, in which the density profile of a quasicrystal
is assumed to be given by a sum of Gaussian functions centered at
the lattice points of a predetermined quasicrystalline
lattice\,\cite{mcarley1994hard}. The width of the Gaussian
function is treated as a variational parameter, which is
optimized to minimize the free energy of the system.
More recently, Jiang and Zhang \cite{jiang2014numerical} proposed the projection method
which embeds the Fourier space of quasicrystal into a
higher-dimensional periodic structure, so that the
quasiperiodic pattern can be recovered by projecting the higher-dimensional reciprocal lattice vectors back to the original
Fourier space through a projection matrix. As a special case,
it can also be used to investigate periodic crystals
by setting the projection matrix as an identity matrix.
From this perspective the projection method provides a
unified computational framework for the study of periodic crystals and quasicrystals.
In this work, we will continue to develop the projection method
by optimizing the shape and size of higher-dimensional periodic cell
which is important to explore potential metastable ordered patterns and
evaluate the free energy with higher accuracy.

The organization of the rest of article is as follows.
In Section \ref{Section: Model}, the coupled-mode Swift-Hohenberg
model with two coupled fields is presented.
We will see that no forementioned artificial assumption is needed
in its analysis with the extension.
In Section \ref{Section: Methods}, we will discuss a
two-mode approximation approach and a computational framework to analyze
the proposed model.
The former mainly focus on the behavior of the model in
the limiting regime, when strong constraints are imposed to the
wave numbers; the latter concerns the general case.
Our numerical method is based on the projection method
\cite{jiang2014numerical}, combined with a variable cell method
optimizing higher-dimensional periodic cells in order to deal with potentially quasicrystalline patterns and other complex phases.
In Section \ref{Section: Results}, phase diagrams for stable
decagonal and dodecagonal quasicrystals in the coupled-mode
Swift-Hohenberg model will be exhibited.
We will also emphasize the important roles of the ratio of characteristic
length scales in pattern formation.
Section \ref{Section: Conclusion} will be devoted to conclusion and a brief discussion.
We will remark that, the numerical framework we develop here can well apply to a large class of more sophisticated models for soft quasicrystals, for example, those employing convolution-type kernels of pairwise correlations \cite{barkan2011stability,rottler2012morphology}.
And our study of the coupled-mode Swift-Hohenberg
model essentially promotes understanding in more general
soft-matter cases.

\section{Coupled-Mode Swift-Hohenberg Model\label{Section: Model}}

We consider the stability of soft quasicrystals in three-component systems,
such as ABC star-shaped terpolymers \cite{hayashida2007polymeric}, ABA'C tetrablock copolymers \cite{zhang2012dodecagonal},
in the Landau theoretical framework.
Quantities of interest are effective densities of each component,
defined by the deviation of spatial densities of monomers from their corresponding averages.
We denote the effective densities of components A, B, and C by $\Phi_A$, $\Phi_B$ and $\Phi_C$ respectively.
By assuming local incompressibility, $\Phi_A+\Phi_B+\Phi_C=0$,
the free energy of the three-component system can be expressed in terms of two order
parameters \cite{dotera2007mean}, $\psi = \Phi_A+\Phi_B=-\Phi_C$ and $\phi= \Phi_A-\Phi_B$,
which should appear in the Landau-type free energy functionals.

Then we construct the minimal model to stabilize the periodic and quasiperiodic patterns.
Followed by the work of Alexander and McTague \cite{alexander1978should},
M{\"u}ller \cite{muller1994model}, and Dotera
\cite{dotera2007mean}, the nonlinear term in the free energy functional containing two-, three-, and four-body
interactions is written as
\begin{equation}
f_{\mathrm{nonlinear}}[\psi, \phi] = \frac{1}{V}\int
\mathrm{d}\mathbf{r}\,[\tau\psi^2
+g_0\psi^3+\psi^4+t\phi^2+t_0\phi^3+\phi^4-g_1\psi^2\phi-g_2\psi\phi^2].
\label{Equation: dotera's original model}
\end{equation}
where $t$, $\tau$, $t_0$, $g_0$, $g_1$ and $g_2$ are parameters
depending on the interaction between components and the
thermodynamic conditions, such as temperature.
In the above free energy functional, the third-order terms play an
important role in stabilizing periodic and quasiperiodic
crystals, because the cubic term is associated with
triad interactions among wave modes.
According to Dotera,
when the component of C dominates the disorder-order transition, the term
$\phi^3$ in \eqref{Equation: dotera's
original model} can be neglected \cite{dotera2007mean}, i.e.~we set $t_0=0$.
The quartic terms are responsible for providing a lower bound for
the free energy. Since only the minimal model is
constructed to study ordered patterns of interest, for simplicity,
the quartic cross terms are omitted.

As is mentioned in the preceding section, the formation of quasicrystals
results from multiple characteristic length scales.
In order to introduce multiple frequency forcing, inspired by the
original work of Swift and Hohenberg
\cite{swift1977hydrodynamic}, later by M{\"u}ller
\cite{muller1994model}, and Lifshitz and Petrich
\cite{lifshitz1997theoretical}, we add two positive-definite
gradient terms into the free energy functional, acting on two
order parameters respectively,
\begin{equation}
\begin{aligned}
f[\psi,\phi]=\frac{1}{V}\Big\{\frac{c}{2}\int &\mathrm{d}\mathbf{r}\,\left[(\nabla^2 +1)\psi\right]^2+\left[(\nabla^2+q^2)\phi\right]^2
\\
&
+\int\mathrm{d}\mathbf{r}\,(\tau\psi^2+g_0\psi^3+\psi^4+t\phi^2+t_0\phi^3 +\phi^4-g_1\psi^2\phi-g_2\psi\phi^2)
\Big\}.
\end{aligned}
\label{Equation: modified_model}
\end{equation}
They act as soft constraints on the
magnitude of wave vectors, or the wave numbers, of the order parameters, with $c>0$ indicating their intensity.
Two critical wave numbers are preferred, $k_c=1$ and $q$ ($q\neq 1$).
The Fourier modes with wave numbers away from them
will increase the free energy in the differential terms and thus are suppressed.
Indeed, if we formally apply Parseval's identity, we find that
\begin{equation}
\int\mathrm{d}\mathbf{r}\,\left[(\nabla^2+1)\psi\right]^2+\left[(\nabla^2+q^2)\phi\right]^2=\frac{1}{(2\pi)^d}\int\mathrm{d}\mathbf{k}\,(-|\mathbf{k}|^2+1)^2|\hat\psi(\mathbf{k})|^2 +(-|\mathbf{k}|^2+q^2)^2|\hat\phi(\mathbf{k})|^2,
\label{Equation: apply parseval's identity to the modified model}
\end{equation}
where $\hat\psi$ (resp.~$\hat\phi$) is the Fourier transform of $\psi$ (resp.~$\phi$) and $d$ is the spatial dimension.
The expression is similar in the case when $\psi$ and $\phi$ can be characterized by Fourier series instead of Fourier transform, simply obtained by replacing the integrations by summations.
As a result, in order to obtain lower energy, the Fourier wave vectors of $\psi$ (resp.~$\phi$) tend to be close to the circle with radius 1 (resp.~$q$) in the Fourier space.
If $c\rightarrow +\infty$, $\hat\psi$ (resp.~$\hat\phi$) should be strictly supported on the circle with radius 1 (resp.~$q$).
This recovers the forementioned assumption that the wave vectors
corresponding to $\psi$ and $\phi$ should favor a single
magnitude in much literature \cite{mermin1985mean,
dotera2007mean}, and makes this model numerically computable.
It should be noted that, there might be only one principal wave number for periodic structures,
which implies that one of the order parameters is degenerate.
For quasicrystals that are a class of multimode patterns and have multiple principal wave numbers,
both of the critical wave numbers should be occupied
\cite{lifshitz1997theoretical, jiang2014LPmodel}.

\section{Methods\label{Section: Methods}}
\subsection{Two-mode approximation method in the limiting regime
$c\rightarrow +\infty$\label{Subsection: Methods for studying asymptotic phase behavior of the model}}
It is helpful to study the phase behavior of the coupled-mode
Swift-Hohenberg model under the limit $c\rightarrow +\infty$.
As is noted above, the analysis performed for various Landau-type
models in much literature is essentially of this type
\cite{chaikin1995principles, mermin1985mean,lifshitz1997theoretical,dotera2007mean}.
In the following, we need this type of asymptotic results to compare with those obtained under finite $c$'s.
Hence we discuss the two-mode approximation approach briefly.

For $c\rightarrow+\infty$, the wave numbers of order parameter $\psi$ (resp.~$\phi$) should be strictly equal to $1$ (resp.~$q$).
We can formally expand them as
\begin{equation}
\psi(\mathbf{x}) = \sum_{\mathbf{k}\in K_\psi\atop|\mathbf{k}|=1} \hat\psi_\mathbf{k}\mathrm{exp}(i\mathbf{k}\cdot\mathbf{x}),\quad \phi(\mathbf{x}) = \sum_{\mathbf{k}\in K_\phi\atop|\mathbf{k}|=q} \hat\phi_\mathbf{k}\mathrm{exp}( i\mathbf{k}\cdot\mathbf{x}).
\label{Equation: spectrum representation of order parameters}
\end{equation}
$K_\psi$ (resp.~$K_\phi$) is the set of spectrum points of $\psi$ (resp.~$\phi$), such that $\mathbf{k}\in K_\psi$ (resp.~$\mathbf{k}\in K_\phi$) if and only if $\mathbf{-k}\in K_\psi$ (resp.~$\mathbf{-k}\in K_\phi$).
$\hat\psi_\mathbf{k}$'s (resp.~$\hat\psi_\mathbf{k}$'s) are complex Fourier coefficients of $\psi$ (resp.~$\phi$) corresponding to the wave vector $\mathbf{k}$.
Since $\psi$ and $\phi$ are real-valued functions, the coefficients should satisfy $\hat\psi_\mathbf{-k} = \overline{\hat\psi_\mathbf{k}}$ and $\hat\phi_\mathbf{-k} = \overline{\hat\phi_\mathbf{k}}$.
Substituting (\ref{Equation: spectrum representation of order
parameters}) into (\ref{Equation: modified_model}), we
immediately get the energy functional as a function of $\hat\psi_\mathbf{k}$'s and $\hat\phi_\mathbf{k}$'s.

Based on the choice of $q$ and the rotational symmetry of interest, we will a priori select several candidates by determining $K_\psi$ and $K_\phi$ as in \cite{dotera2007mean}.
They will be specified below right before each numerical simulation.
For simplicity, we also choose $|\hat\psi_\mathbf{k}|$ (resp.~$|\hat\phi_\mathbf{k}|$) to be identical for all $\mathbf{k}\in K_\psi$ (resp.~$\mathbf{k}\in K_\phi$).
Then we use the common optimization method, e.g.,~the steepest descent method, to minimize the free energy $f$ with respect to $\hat\psi_\mathbf{k}$'s and $\hat\phi_\mathbf{k}$'s for each candidate.
Note that $\hat\psi_\mathbf{k}$'s and $\hat\phi_\mathbf{k}$'s are complex numbers, which implies we should not only optimize the amplitude of plane waves, but also their phase angles.
Finally, we choose the configuration with the lowest energy as the ground state.
Repeat the above procedure for all the parameter points of interest, we will obtain the phase diagram in the limit $c\rightarrow +\infty$.

\subsection{Direct minimization approach with the variable cell method\label{Subsection: The strategy of optimizing the computational domain}}
In order to study the coupled-mode Swift-Hohenberg model for general $c$'s, we propose the following numerical method to directly minimize the free energy functional (\ref{Equation: modified_model}).

We basically employ the steepest descent method in the minimization: introduce an auxiliary variable $t$ and solve the following relaxation equations from some initial configurations $\psi(\mathbf{x},0)$ and $\phi(\mathbf{x},0)$ until the equilibrium is reached.
\begin{equation}
\begin{split}
\frac{\partial\psi}{\partial t}=-\frac{\delta f}{\delta\psi}=&-c(\nabla^2+1)^2\psi-(2\tau\psi+3g_0\psi^2+4\psi^3)+2g_1\psi\phi+g_2\phi^2\\
\frac{\partial\phi}{\partial t}=-\frac{\delta
f}{\delta\phi}=&-c(\nabla^2+q^2)^2\phi-(2t\phi+3t_0 \phi^2+
4\phi^3) +2g_2\psi\phi+g_1\psi^2.
\end{split}
\end{equation}
These equations are solved in a pseudo-spectral approach.
Take Fourier transform on both sides and we find
\begin{equation}
\begin{split}
\frac{\partial\hat{\psi}}{\partial t}=&-c(-|\mathbf{k}|^2+1)^2\hat{\psi}-(2\tau\psi+3g_0\psi^2+4\psi^3)^{{}_{\displaystyle\hat{}}}+2g_1(\psi\phi)^{{}_{\displaystyle\hat{}}}+g_2(\phi^2)^{{}_{\displaystyle\hat{}}}\\
\frac{\partial\hat{\phi}}{\partial t}=&-c(-|\mathbf{k}|^2+q^2)^2\hat{\phi}-(2t\phi+3t_0\phi^2+4\phi^3)^{{}_{\displaystyle\hat{}}}+ 2g_2(\psi\phi)^{{}_{\displaystyle\hat{}}} +g_1(\psi^2)^{{}_{\displaystyle\hat{}}}.
\end{split}
\label{Equation: steepest descent method in Fourier space}
\end{equation}
The differential terms can be easily evaluated by pointwise
multiplication in the Fourier space, while for the polynomial terms, we first perform inverse Fourier transform to obtain $\psi$ and $\phi$. Then we compute them in the real space and perform Fourier transform to get the desired form in the Fourier space.

We note that the pseudo-spectral method is suitable for this problem because all the periodic and quasiperiodic orders $\psi$ and $\phi$ will have discrete Fourier spectrum.
On the other hand, it requires careful discretization of the Fourier space so that all the spectrum points can be well captured.
The periodic orders are discretized in an ordinary way since their spectrum points locate on a periodic lattice in the Fourier space.
To deal with the potentially quasiperiodic states, we employ the projection method \cite{jiang2014numerical}.
For 2-dimensional quasicrystals,
more than two basis vectors are used in the Fourier space to represent the discrete spectrum of quasicrystalline patterns.
To be more precise, we need $n>2$ basis vectors in $\mathbb{R}^2$
that are linearly independent in the field of rational number
$\mathbb{Q}$, namely, $\mathbf{e}_1, \mathbf{e}_2, \cdots,
\mathbf{e}_n\in\mathbb{R}^2$. Then all the discretization points in the Fourier space are encoded as
\begin{equation}
\mathbf{k} = \sum_{i=1}^n a_i\mathbf{e}_i,\quad a_i\in\{-N+1,\cdots,N\}.
\label{Equation: discretization point in projection method}
\end{equation}
The linear independence guarantees that different $(a_1,\cdots,a_n)$'s will give different $\mathbf{k}$'s.
The choice of $n$ and $\mathbf{e}_i$ closely relies on the orders of interest, especially their symmetry.
In particular, $n = 4$ for decagonal and dodecagonal quasicrystals. The basis vectors are initially chosen as
\begin{equation}
\mathbf{e}_i = \left(\cos\frac{2j\pi}{m}, ~ \sin\frac{2j\pi}{m}\right),\quad j = 1,2,3,4,
\end{equation}
where $m = 10$ (resp.~$m=12$) in the decagonal (resp.~dodecagonal) case.
We will discuss the approach of optimizing computational cell, or
equivalently adjusting basis vectors later.
Then all the computations will well resemble those in computing a periodic order using spectral method in $n$-dimensional space.
Readers are referred to \cite{jiang2014numerical} for more discussions on technical details.
We only note that the computational cost is considerably large in such higher-dimensional computations.
For example, for decagonal and dodecagonal quasicrystals, we use $O(N^4)$ spectrum points, where $N$ is the number of discretization points on each dimension given in (\ref{Equation: discretization point in projection method}).
Hence the cost in each step of iteration in the steepest descent method grows as $O(N^4\log N)$ since we heavily use FFT.

It follows that Equations \eqref{Equation: steepest descent method in Fourier space} becomes a group of ODEs for the Fourier coefficients on the given spectrum points.
Due to numerical stiffness, semi-implicit scheme is employed in the iteration to avoid extremely small time step.
For simplicity, we use fixed time step.
Initial value of the iteration is given by approximants of potential ground states, such as some quasicrystalline orders of interest or some periodic orders whose Fourier spectrum consists of plane waves with magnitude $1$ and $q$ only.
We will specify these initial values later right before each numerical experiment.

As is discussed above, we choose the basis $\mathbf{e}_i$ (and thus all the spectrum points of $\psi$ and $\phi$) in the numerical method.
However, it is often difficult to determine a priori where the spectrum of the ground state should locate.
The choice of basis becomes an artificial restriction in the computations.
Hence, we are going to improve the minimization method with a strategy of optimizing the computational domain, which is named variable cell method.

Along with (\ref{Equation: steepest descent method in Fourier space}), we also apply the steepest descent method to optimize the basis vectors, i.e.,
\begin{equation}
\frac{\partial \mathbf{e}_i}{\partial t}=-\lambda \frac{\partial f}{\partial\mathbf{e}_i},\quad i = 1,2,\cdots,n,
\label{Equation: steepest descent method in Fourier space for basis}
\end{equation}
where $\lambda$ is a positive constant determined empirically to make the evolution numerically stable and efficient.
Explicit scheme and fixed time step are employed.
Since $f$ is originally not a function of the basis vectors $\mathbf{e}_i$'s,
we note that they actually come into the differential term of the free energy functional.
To see this, we can expand $\psi$ and $\phi$ as in (\ref{Equation: spectrum representation of order parameters}).
Then the free energy becomes
\begin{equation}
f[\psi,\phi] = c'\sum_{\mathbf{k}}\left[(-|\mathbf{k}|^2+1)^2|\hat\psi_{\mathbf{k}}|^2 +(-|\mathbf{k}|^2+q^2)^2|\hat\phi_{\mathbf{k}}|^2\right]+g(\hat\psi,\hat\phi),
\label{Equation: express f as a function of basis}
\end{equation}
where $c'$ is a positive number depending on $c$ and the constant coming out from the integral.
$\hat\psi$ (resp.~$\hat\phi$) is a long vector whose components are all $\hat\psi_{\mathbf{k}}$'s (resp.~$\hat\phi_{\mathbf{k}}$'s).
$g(\hat\psi,\hat\phi)$ is the energy contributed by the polynomial terms in (\ref{Equation: modified_model}), which does not depend on the position of Fourier spectral points (and thus the basis vectors) but only their coefficients.
One then expands $\mathbf{k}$ in the above formula by using \eqref{Equation: discretization point in projection method}.
It is easy to see that the free energy becomes a function of $\hat\psi$, $\hat\phi$ and $\mathbf{e}_i$'s.
Hence, \eqref{Equation: steepest descent method in Fourier space for basis} is well-defined.

There is still a special constraint in optimizing basis vectors
in the quasiperiodic orders, when the number of basis vectors is
larger than the dimension of the Fourier space: the basis vectors
should be linearly independent in the field of rational numbers
$\mathbb{Q}$.
Otherwise, the expression of the energy contributed by the polynomial terms (the second term in \eqref{Equation: express f as a function of basis}) will be changed.
Once the basis vectors start to evolve with (\ref{Equation: steepest descent method in Fourier space for basis}), it is difficult to preserve or check this property.
However, a weaker condition is helpful: the basis vectors should not be parallel with each other.
This condition can be violated in the numerical sense.
For example, when one starts minimization with a quasiperiodic configuration using four basis vectors under some parameters, where a two-dimensional periodic order is the actual ground state, the four basis vectors may tend to ``mimic" the spectrum of the two-dimensional periodic order.
Some basis vectors will thus become almost parallel to each other and the problem becomes ill-conditioned.
To formulate this criterion, we let $\mathbf{e}_i = (e_i^1,e_i^2)$, $i=1,2,\dots,n$, and require that
\begin{equation}
\left|e_i^1e_j^2 - e_j^1e_i^2\right| \geqslant \epsilon,\quad \forall\,i\not= j,
\label{Equation: condition for well-posed basis vectors}
\end{equation}
where $\epsilon$ is a small positive number whose typical value is chosen as $0.05$.
In every step of steepest descent iteration, we will check if (\ref{Equation: condition for well-posed basis vectors}) is satisfied.
If not, the optimization has to be terminated.
In computing periodic orders where the number of basis vectors is equal to the dimension of the Fourier space, the same criterion may also apply although it is not necessary.

Based on the discussions above, we conclude the direct
minimization algorithm by sketching the numerical recipes
\begin{description}
  \item[Step 1] Given parameters $c$, $q$, $t$, $t_0$, $\tau$, $g_0$,
	  $g_1$ and $g_2$, choose the basis vectors and the initial configurations of $\psi$ and $\phi$.
  \item[Step 2] Freeze the basis vectors and minimize (\ref{Equation: modified_model}) with respect to $\psi$ and $\phi$ by (\ref{Equation: steepest descent method in Fourier space}). The iteration is terminated when the right-hand side of (\ref{Equation: steepest descent method in Fourier space}) is sufficiently small.
  \item[Step 3] Minimize (\ref{Equation: modified_model}) with respect to $\psi$, $\phi$ and basis vectors simultaneously by (\ref{Equation: steepest descent method in Fourier space}) and (\ref{Equation: steepest descent method in Fourier space for basis}). Condition (\ref{Equation: condition for well-posed basis vectors}) is checked after each step of iteration. The minimization is terminated when the right-hand sides of (\ref{Equation: steepest descent method in Fourier space}) and (\ref{Equation: steepest descent method in Fourier space for basis}) are both sufficiently small, which is a good ending, or when (\ref{Equation: condition for well-posed basis vectors}) is violated, which is a bad ending. In the latter case, a new initial configuration has to be chosen.
  \item[Step 4] Repeat the above process with various initial configurations and find out the ground state with the lowest free energy.
\end{description}
By repeating the above steps with different choice of parameters, we can obtain phase diagrams of interest. We also remark that Step 2 helps improve the efficiency of the algorithm, since the initial guess of basis vectors can be far away from the final configuration, which may lead to long iterations if we optimize order parameters and basis vectors simultaneously from the beginning.

In principle, the developed numerical methods can be applied to
compute two- and three-dimensional quasiperiodic patterns. The
dimension $n$ of computational space in the projection method
depends on the symmetries of considered quasicrystals
\cite{jiang2014numerical}. For example, for three-dimensional
icosahedral quasicrystals, the projection method should be
performed in six-dimensional space, which results in huge
computational cost. Hence, we restrict the
current study to two-dimensional quasiperiodic patterns whose point group
symmetries can be embedded into four-dimensional periodic lattice,
such as quasicrystalline orders with 10- or 12-fold symmetry.

\section{Numerical Results\label{Section: Results}}
\subsection{$t-\tau$ phase diagrams in the limiting regime $c\rightarrow +\infty$\label{Subsection: t-tau phase diagrams by the asymptotic study}}
\begin{figure}[t]
\centering
\includegraphics[width=2.8cm]{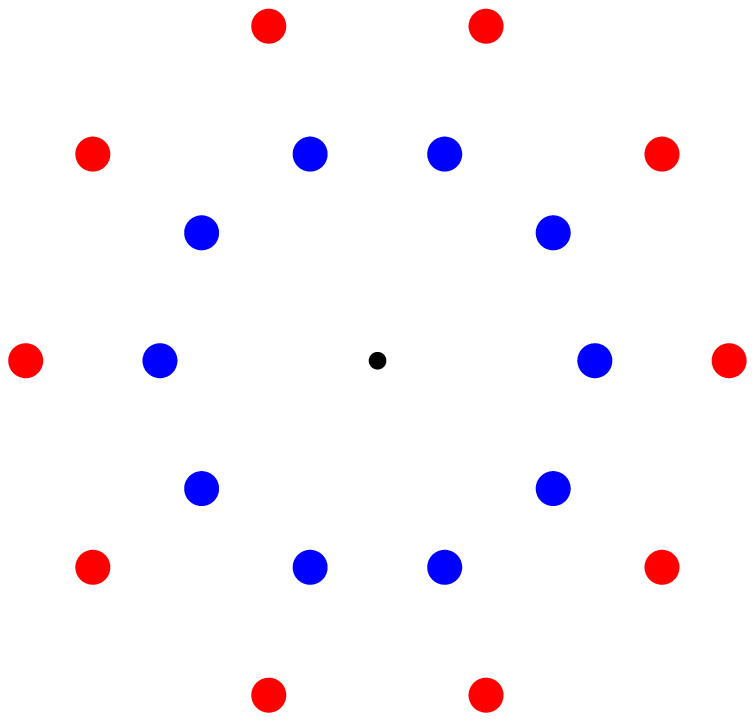}
\includegraphics[width=2.8cm]{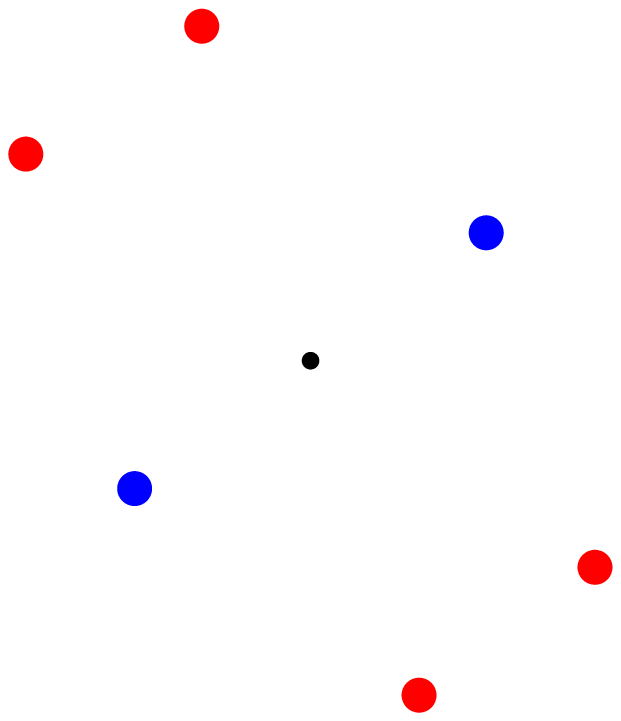}
\includegraphics[width=2.8cm]{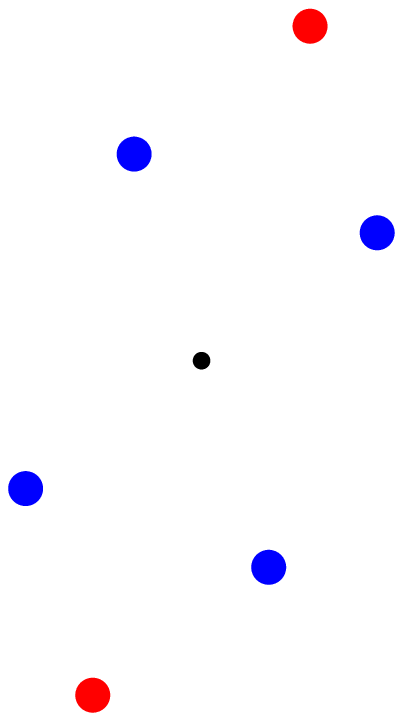}
\\
\includegraphics[width=2.8cm]{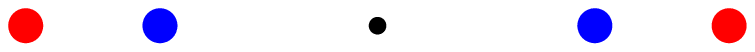}
\includegraphics[width=2.8cm]{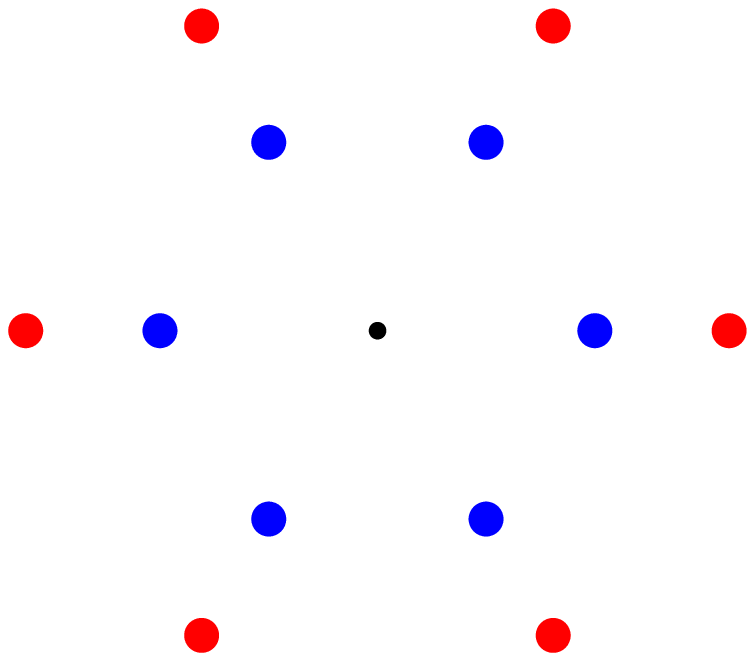}
\includegraphics[width=2.8cm]{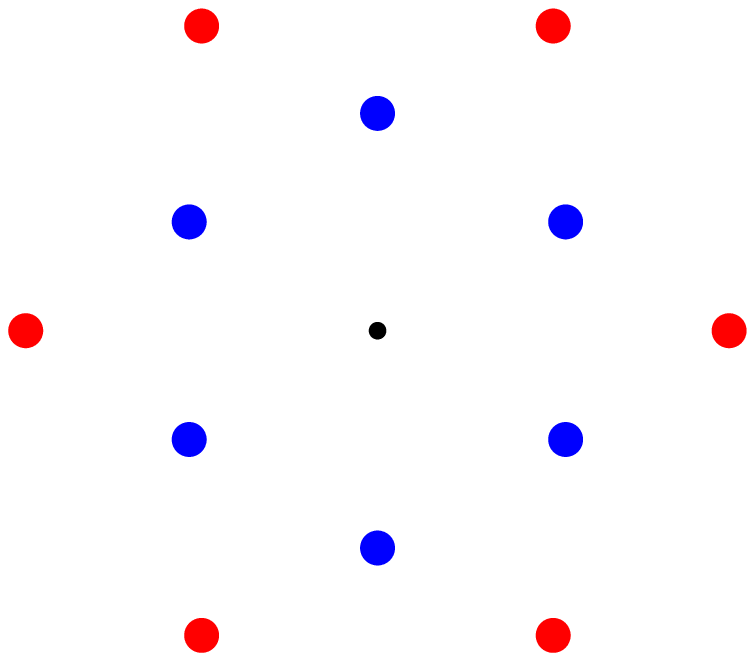}
\caption{Fourier spectrum of candidates in the decagonal case: $q = 2\cos\frac{\pi}{5}$, $t_0 = 0$, $g_0 = 0.2$, $g_1=2.2$ and $g_2=2.2$. The black dot in each small figure represents the origin of the Fourier space.
The blue and red dots surrounding the origin are end points of the wave vectors in $K_\psi$ and $K_\phi$ respectively.}
\label{Figure: D_candidates}
\end{figure}
\begin{figure}[t]
\centering
\includegraphics[width=2.8cm]{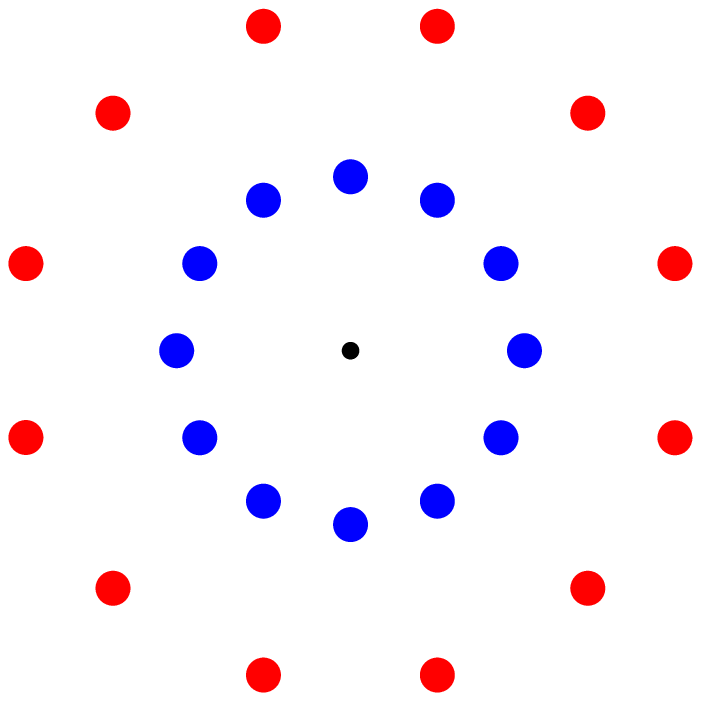}
\includegraphics[width=2.8cm]{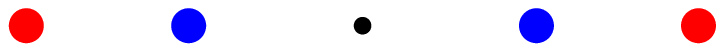}
\includegraphics[width=2.8cm]{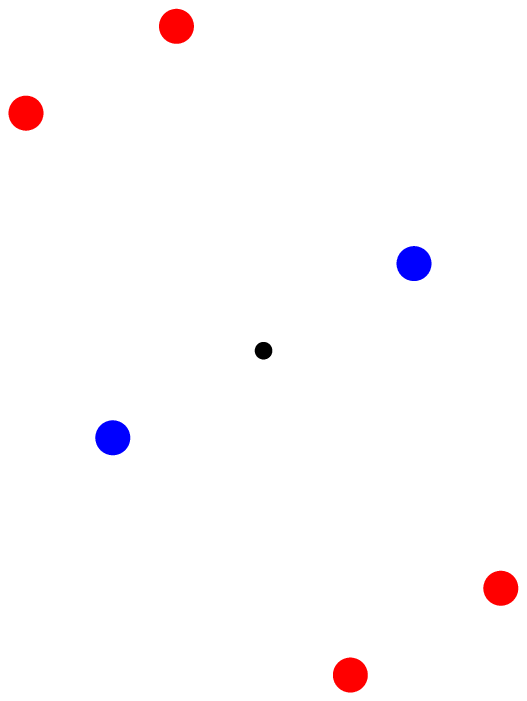}
\includegraphics[width=2.8cm]{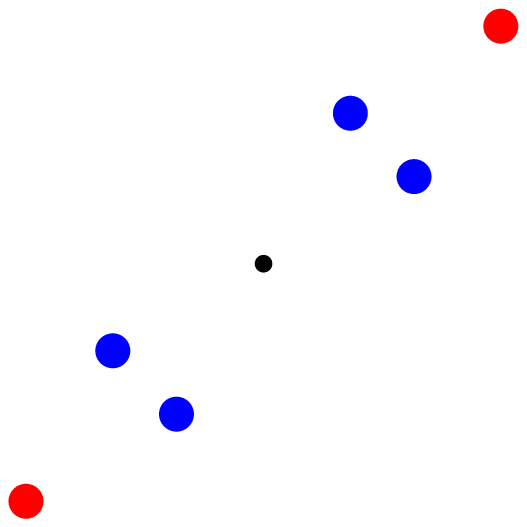}\\
\includegraphics[width=2.8cm]{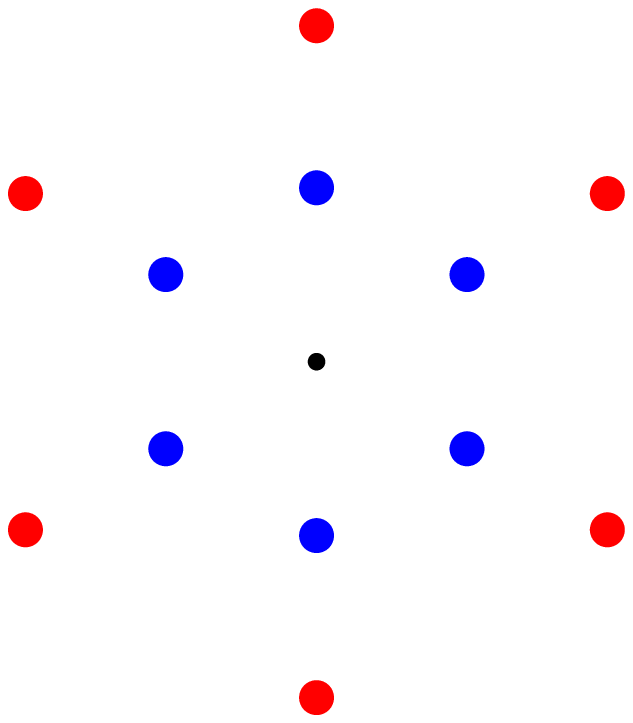}
\includegraphics[width=2.8cm]{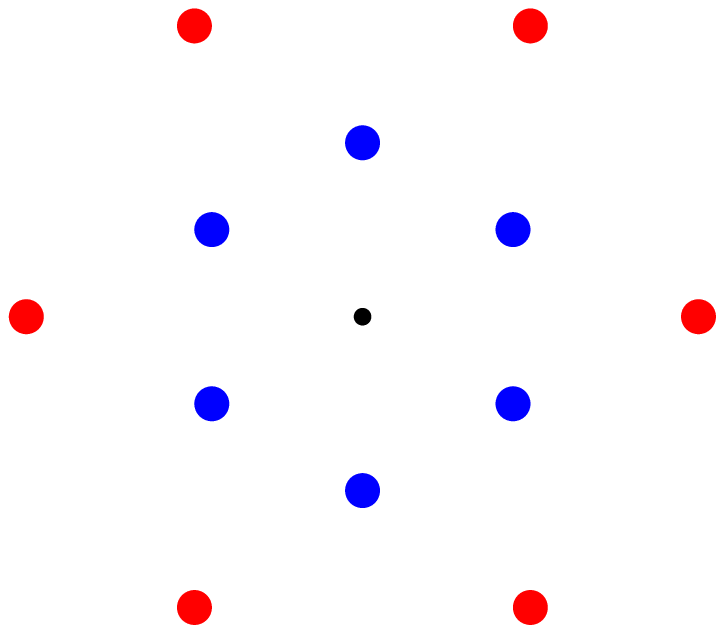}
\includegraphics[width=2.8cm]{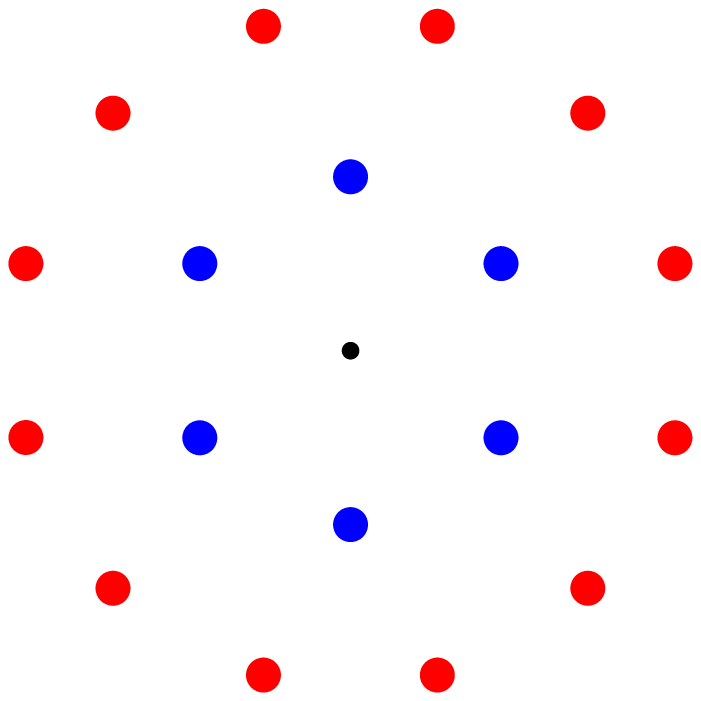}
\includegraphics[width=2.8cm]{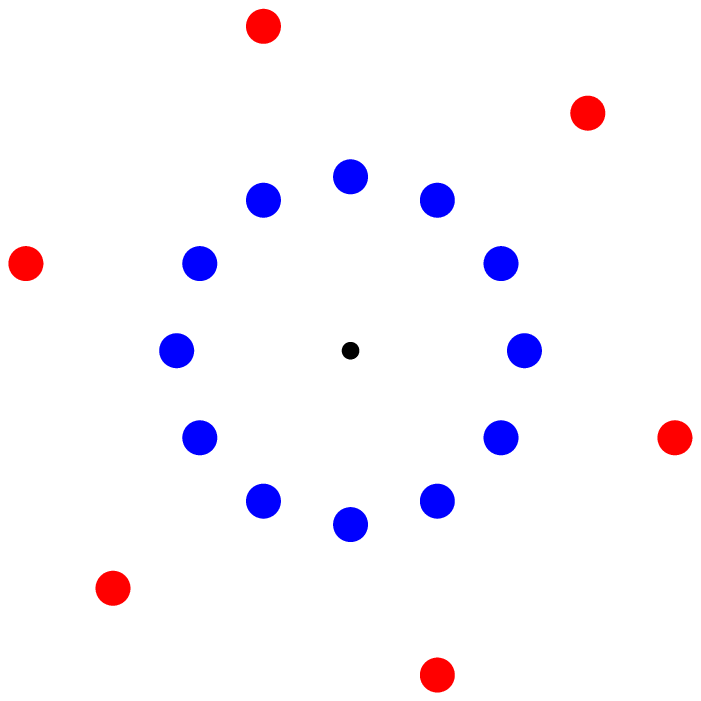}
\caption{Fourier spectrum of candidates in the dodecagonal case: $q = 2\cos\frac{\pi}{12}$, $t_0=0$, $g_0=0.8$, $g_1 = 2.2$ and $g_2=0.2$. The black, blue and red dots have the same meanings as in FIG.~\ref{Figure: D_candidates}.}
\label{Figure: DD_candidates}
\end{figure}
We will focus on searching for decagonal and dodecagonal quasicrystalline orders under the limit $c\rightarrow+\infty$ in this subsection.
Due to the geometric features of the desired patterns, we will
choose $q = 2\cos\frac{\pi}{5}$ and $q = 2\cos\frac{\pi}{12}$ for
the decagonal case and dodecagonal case, respectively, as in
\cite{dotera2012toward, jiang2014LPmodel}.
We also let $t_0 = 0$, $g_0 = 0.2$, $g_1=2.2$ and $g_2=2.2$ for the
decagonal case, and $t_0 = 0$, $g_0=0.8$, $g_1 = 2.2$ and $g_2=0.2$ for the dodecagonal case.
We leave $t$ and $\tau$ as free parameters forming the phase space.
The above settings are the same as in \cite{dotera2007mean}.

In the decagonal case ($q=2\cos(\pi/5)$) and the dodecagonal case
($q=2\cos(\pi/12)$), we select
$K_\psi$'s and $K_\phi$'s of the candidates as in
FIG.~\ref{Figure: D_candidates} and FIG.~\ref{Figure:
DD_candidates} respectively.
The black dot in each small figure represents the origin of the Fourier space.
The blue (resp.~red) dots surrounding the origin are end points of wave vectors in $K_\psi$ (resp.~$K_\phi$).
These candidates are chosen either because large numbers of
triangles can be formed in the Fourier space using their wave
vectors \cite{lifshitz1997theoretical} which can reduce the free
energy value, or because their Fourier spectrum is similar with
that of some widely observed orders
in structured soft materials, such as lamellae phase and lamellae
phase with alternating beads in block polymer
systems\cite{tang2004morphology,matsushita2007creation,
xu2013strategy}.
We also remark that the first candidates in FIG.~\ref{Figure: D_candidates} and FIG.~\ref{Figure: DD_candidates} enjoy decagonal and dodecagonal rotational symmetry respectively.

\begin{figure}[!htbp]
\centering
\subfigure[Decagonal case]
{\label{Subfigure: D_phase_diag}
\includegraphics[scale = 0.4]{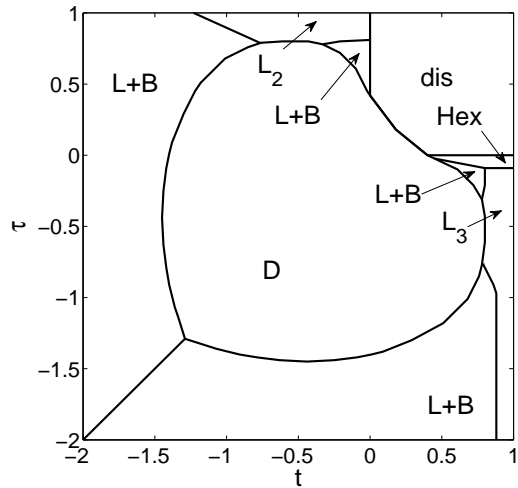}}\\
\subfigure[Dodecagonal case]{\label{Subfigure:
DD_phase_diag}\includegraphics[scale = 0.4]{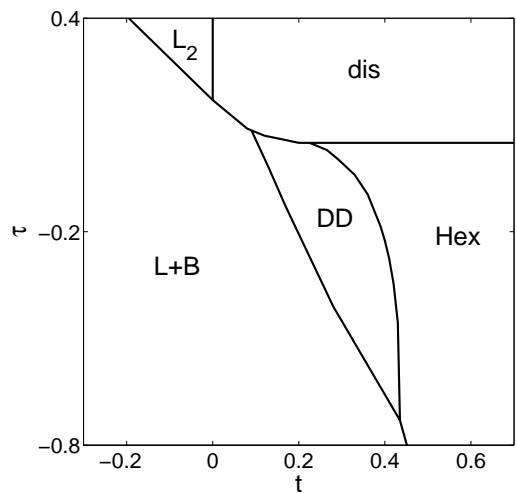}}
\caption{Phase diagrams in $t-\tau$ plane under the limit
$c\rightarrow +\infty$: (a) decagonal case: $q =
2\cos\frac{\pi}{5}$, $t_0=0$, $g_0=0.2$, $g_1 = 2.2$, $g_2=2.2$;
(b) dodecagonal case: $q = 2\cos\frac{\pi}{12}$, $t_0=0$, $g_0=0.8$, $g_1 = 2.2$, $g_2=0.2$. The phases are named after their real-space morphology. \emph{D} and \emph{DD} are decagonal and dodecagonal quasicrystals respectively. \emph{L+B} is the lamellae phase with alternating beads. \emph{L}$_2$ and \emph{L}$_3$ are two types of lamellae phases. \emph{Hex} is a hexagonal phase. \emph{Dis} is the uniform, isotropic disordered phase.}
\label{Figure: D DD phase diag asymptotic}
\end{figure}

With the two-mode approximation method discussed in Section \ref{Subsection: Methods for studying asymptotic phase behavior of the model}, we can obtain phase diagram in the $t-\tau$ plane for the decagonal and dodecagonal case in the limit $c \rightarrow +\infty$, shown in FIG.~\ref{Figure: D DD phase diag asymptotic}.
Note that only one $q$, as specified above, is used throughout each individual phase diagram.
The phases are named according to their real-space morphologies.
To be more precise, we first recover the effective densities $\Phi_A$, $\Phi_B$ and $\Phi_C$ from $\psi$ and $\phi$ through
\begin{equation}
\Phi_A = \frac{1}{2}(\psi+\phi),\quad \Phi_B = \frac{1}{2}(\psi-\phi),\quad \Phi_C = -\psi.
\label{Equation: transformation from order parameter fields to effective density fields}
\end{equation}
Then the morphology of phases are determined by plotting dominant regions of the three components; the dominant region of a component is defined to be the region in which its effective density is the highest among the three \cite{tang2004morphology}.
Readers are referred to Section \ref{Subsection: t-tau phase diagrams by the direct minimization with the variable cell method} for illustrations of this transformation.

The phase diagram for the decagonal case (FIG.~\ref{Subfigure: D_phase_diag}) is largely symmetric with respect to the diagonal $t=\tau$.
The decagonal quasicrystal, named \emph{D}, is stable in a heart-shaped region around the critical temperatures of the two order parameters, i.e.~$(t,\tau) = (0,0)$.
Some periodic orders are stable in the regions around.
\emph{L+B} is short for lamellae phase with alternating beads \cite{tang2004morphology}.
It is stable in four separate regions.
\emph{L}$_3$ and \emph{L}$_2$ are two lamellae phases.
In \emph{L}$_3$, all of the three components have their own dominant regions and they form lamellae phase.
In \emph{L}$_2$, two of the components form lamellae phase, while the third component fully blends with them.
\emph{Hex} is a hexagonal phase, where one component is enriched
in a hexagonal lattice and the other two components coexist in
the remaining matrix.
\emph{L}$_2$ and \emph{Hex} can be treat as degenerate phases, since one or more components have no dominant regions in the real space.
They occur when $t$ or $\tau$ is sufficiently positive so that one order parameter has to vanish.
\emph{Dis}, short for disordered phase, is a uniform and isotropic phase where $\psi = \phi = 0$.

In FIG.~\ref{Subfigure: DD_phase_diag}, a small region of the dodecagonal quasicrystal (\emph{DD}) is found close to $(t,\tau) = (0,0)$.
However, the region is entirely in $t>0$ half-plane.
This implies that under the current parameters, the dodecagonal quasicrystalline order of $\phi$ is completely induced by $\psi$ through the coupling effects.
\emph{L+B}, \emph{Hex} and \emph{L}$_2$ are also found in the phase diagram, with the same morphology as above.

We remark that our result is qualitatively similar with Dotera's
\cite{dotera2007mean} in the decagonal case, but it is quite
different in the dodecagonal case. The reasons are threefold:
firstly we optimize all the Fourier coefficients independently
rather than assuming strong rotational symmetry of the
coefficients; secondly, we only put orders corresponding to the
same parameter $q$ into comparison; lastly, we do not consider more
complex orders such as the Archimedean tiling $(3.3.4.3.4)$.

\subsection{$t-\tau$ phase diagrams by the direct minimization with the variable cell method
\label{Subsection: t-tau phase diagrams by the direct minimization with the variable cell method}}
\begin{figure}[!htbp]
\centering
\includegraphics[width=2.8cm]{candidates_Vectors_D_stdD.eps}
\includegraphics[width=2.8cm]{candidates_Vectors_D_i2o4.eps}
\includegraphics[width=2.8cm]{candidates_Vectors_D_i4o2.eps}
\includegraphics[width=2.8cm]{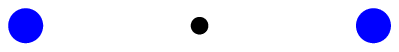}
\includegraphics[width=2.8cm]{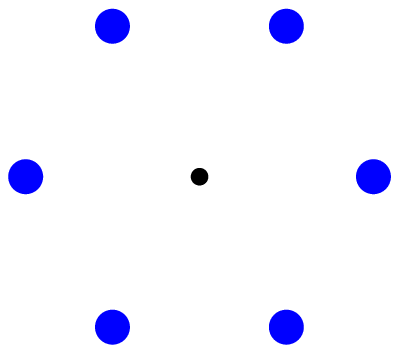}
\caption{Fourier spectrum of initial configurations in the
decagonal case: $q = 2\cos\frac{\pi}{5}$, $t_0 = 0$, $g_0 = 0.2$, $g_1=2.2$ and $g_2=2.2$. The black, blue and red dots have the same meanings as in FIG.~\ref{Figure: D_candidates}.}
\label{Figure: D_initial configurations}
\end{figure}

\begin{figure}[!htbp]
\centering
\includegraphics[width=2.8cm]{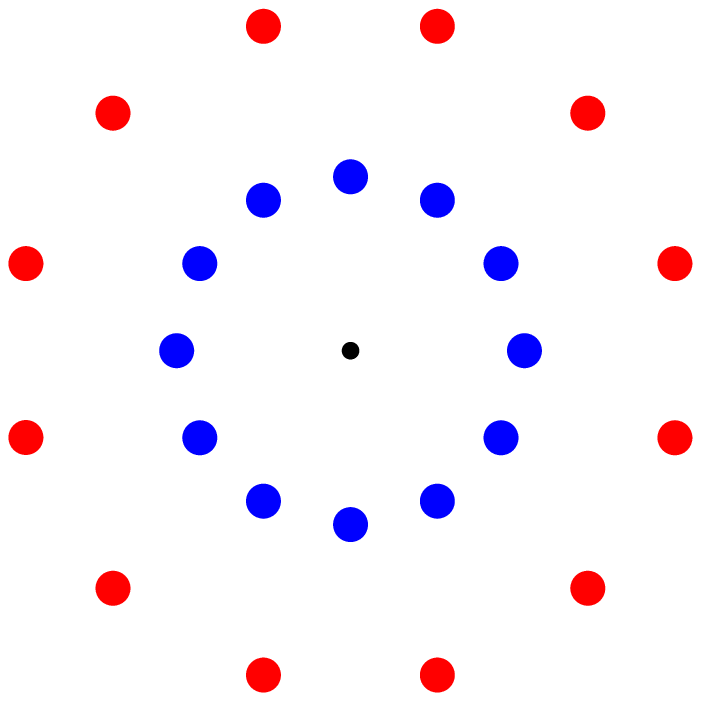}
\includegraphics[width=2.8cm]{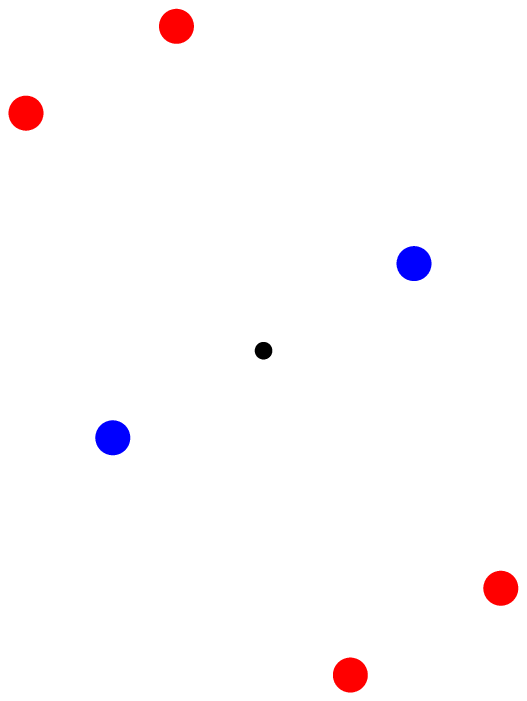}
\includegraphics[width=2.8cm]{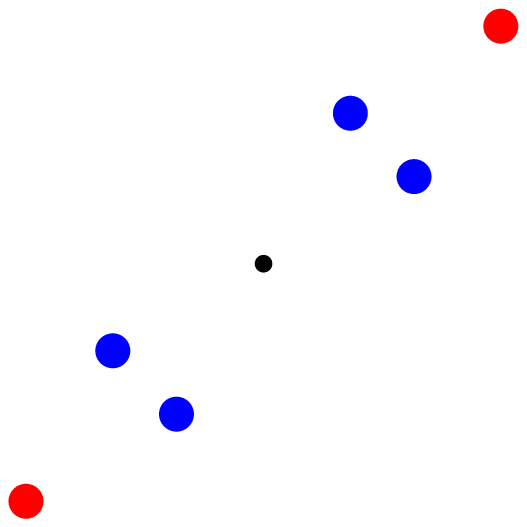}
\includegraphics[width=2.8cm]{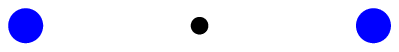}
\includegraphics[width=2.8cm]{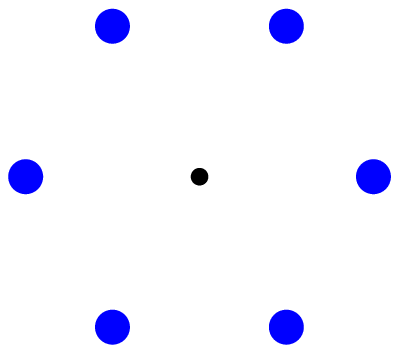}
\caption{Fourier spectrum of initial configurations in the
dodecagonal case: $q = 2\cos\frac{\pi}{12}$, $t_0=0$, $g_0=0.8$, $g_1 = 2.2$ and $g_2=0.2$. The black, blue and red dots have the same meanings as in FIG.~\ref{Figure: D_candidates}.}
\label{Figure: DD_initial configurations}
\end{figure}

We set $c=80$ in the following simulations as an example to
present the phase behavior under finite $c$'s.
The parameters are set as the same as before, namely, $q =
2\cos\frac{\pi}{5}$, $t_0 = 0$, $g_0=0.2$, $g_1 = 2.2$ and
$g_2=2.2$ in the decagonal case, and $q = 2\cos\frac{\pi}{12}$,
$t_0=0$, $g_0=0.8$, $g_1 = 2.2$ and $g_2=0.2$ in the dodecagonal case.
In order to obtain the phase diagrams, we need several initial configurations of the spectrum points.
These initial configurations are selected based on the ground states observed in the preceding phase diagrams in the limiting case, i.e., FIG.~\ref{Figure: D DD phase diag asymptotic}.
One reason is that we believe the phase behavior in the case of $c = 80$ is qualitatively close to that in the limiting case $c\rightarrow+\infty$.
Also, the variable cell method can find the optimal configuration of
spectrum points by itself, since the basis vectors are allowed to vary in the minimization.
Hence, there is no need to choose as many initial candidates as in the limiting case.
We represent the initial configurations again by specifying their Fourier spectrum, i.e.,~$K_\psi$ and $K_\phi$ in (\ref{Equation: spectrum representation of order parameters}).
See FIG.~\ref{Figure: D_initial configurations} for the decagonal case and FIG.~\ref{Figure: DD_initial configurations} for the dodecagonal case.
The black, blue and red dots in these figures have the same meanings as in FIG.~\ref{Figure: D_candidates} and FIG.~\ref{Figure: DD_candidates}.
The first candidate in FIG.~\ref{Figure: D_initial
configurations} (resp.~FIG.~\ref{Figure: DD_initial
configurations}) stands for decagonal (resp.~dodecagonal)
quasicrystal with perfect symmetry. We apply the projection method in the minimization and use four basis vectors.
The other initial candidates are periodic orders in two dimensional space.


\begin{figure}[!htbp]
\centering
\subfigure[Decagonal case]{\label{Subfigure: D_phase_diag c 80}\includegraphics[scale = 0.4]{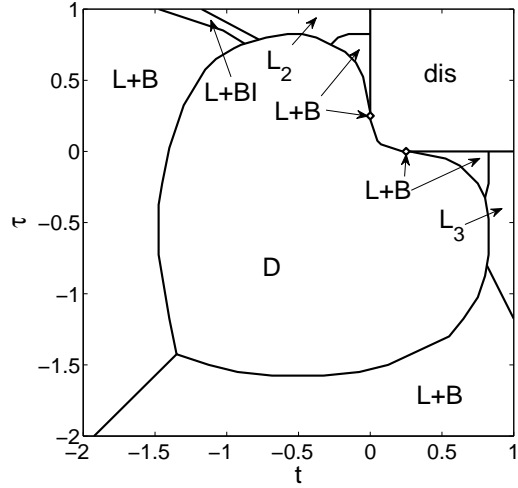}}\\
\subfigure[Dodecagonal case]{\label{Subfigure: DD_phase_diag c 80}\includegraphics[scale = 0.4]{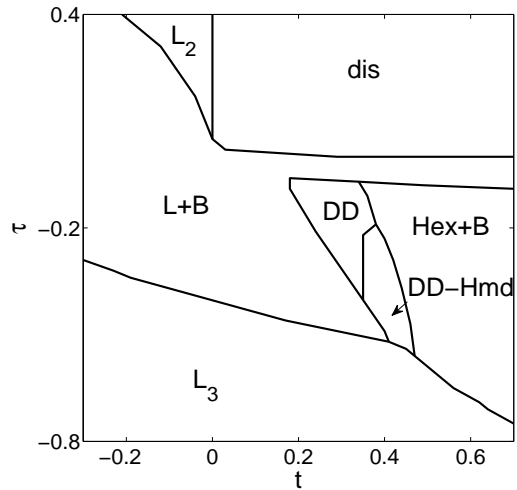}}
\caption{Phase diagrams in $t-\tau$ plane for $c=80$: (a)
decagonal case: $q = 2\cos\frac{\pi}{5}$, $t_0 = 0$, $g_0=0.2$,
$g_1 = 2.2$, $g_2=2.2$; (b) dodecagonal case: $q =
2\cos\frac{\pi}{12}$, $t_0=0$, $g_0=0.8$, $g_1 = 2.2$, $g_2=0.2$. The phases are named after their real-space morphology. \emph{L+BI} is the lamellae phase with beads at the interface, viewed as a transition phase between \emph{L+B} and \emph{L}$_2$. \emph{Hex+B} is the hexagonal phase with beads. \emph{DD-Hmd} is short for dodecagonal quasicrystalline phase with hexagonal modulation, which can be treated as a superposition of a perfect quasiperiodic order with dodecagonal symmetry and a periodic order with hexagonal symmetry, and also a transition phase between the two. Other names of the phases have the same meanings as in FIG.~\ref{Figure: D DD phase diag asymptotic}.}
\label{Figure: D DD phase diag c 80}
\end{figure}
\begin{figure}[!htbp]
\centering
\subfigure[\emph{D} phase]{\label{Subfigure: D D real-morphology}\includegraphics[scale = 0.32, bb = 100 280 500 580]{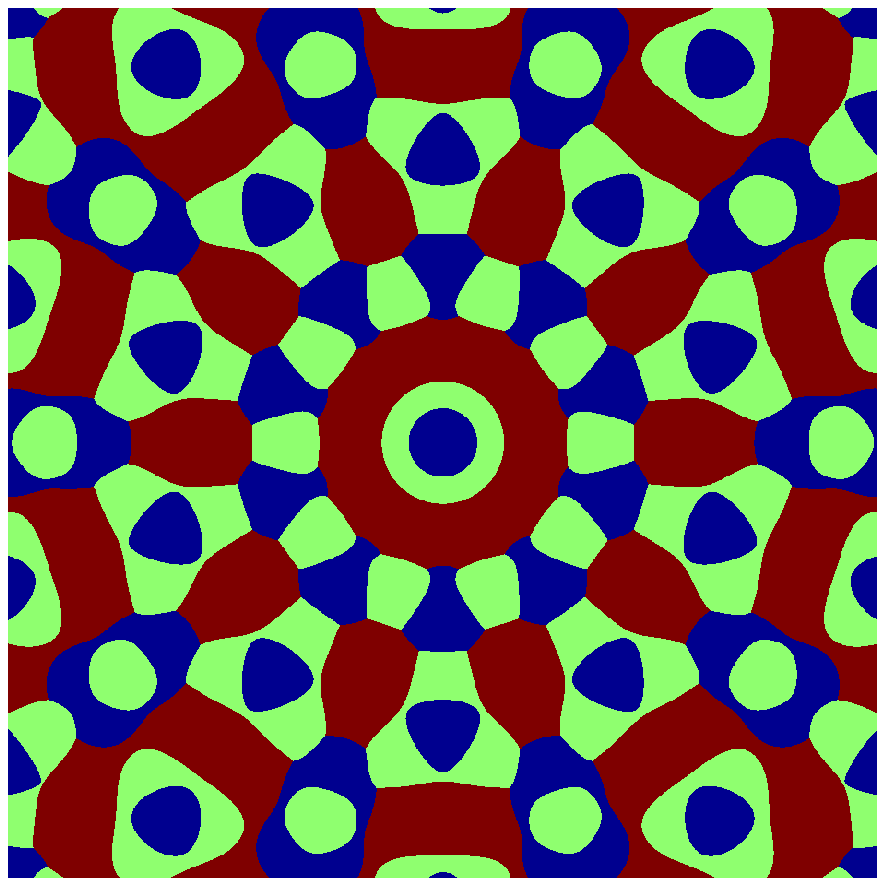}}
\subfigure[Order parameter $\psi$]{\label{Subfigure: D psi}\includegraphics[scale = 0.32, bb = 100 280 500 580]{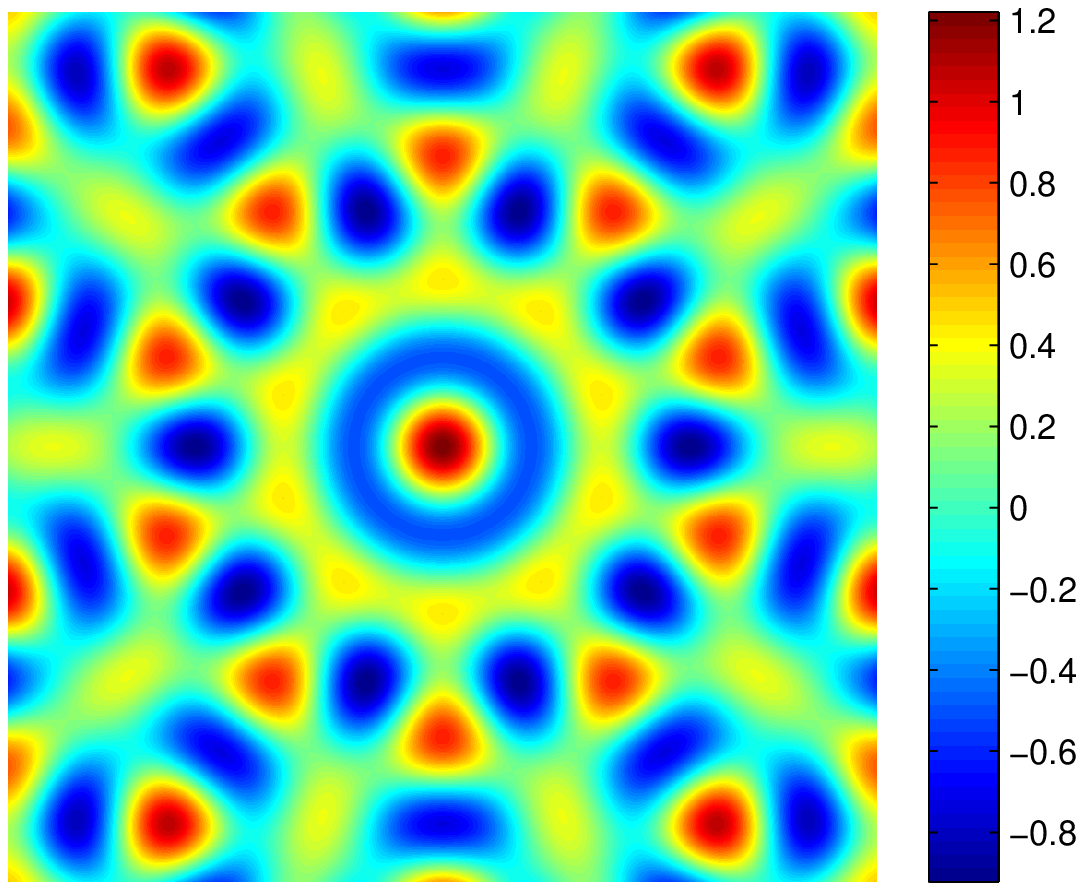}}
\subfigure[Order parameter $\phi$]{\label{Subfigure: D phi}\includegraphics[scale = 0.32, bb = 100 280 500 580]{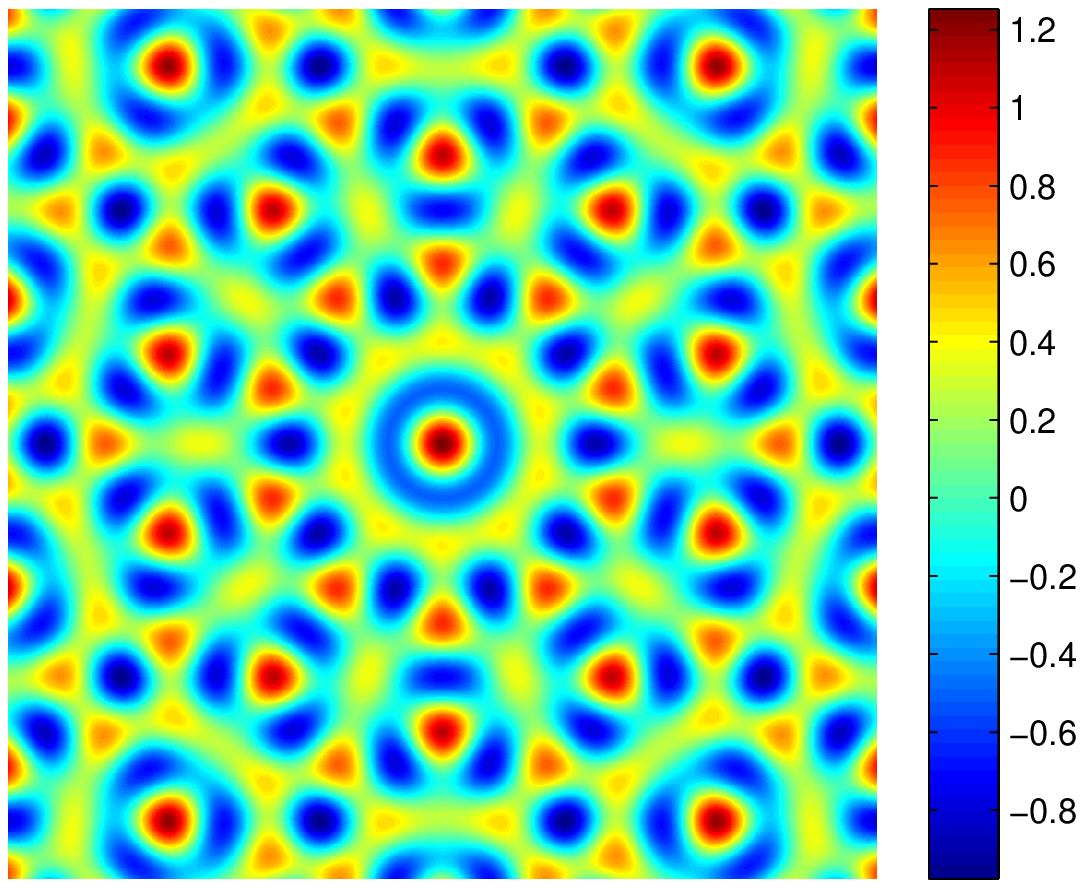}}\\
\subfigure[Effective density of A, $\Phi_A$]{\label{Subfigure: D phi_a}\includegraphics[scale = 0.32, bb = 100 280 500 580]{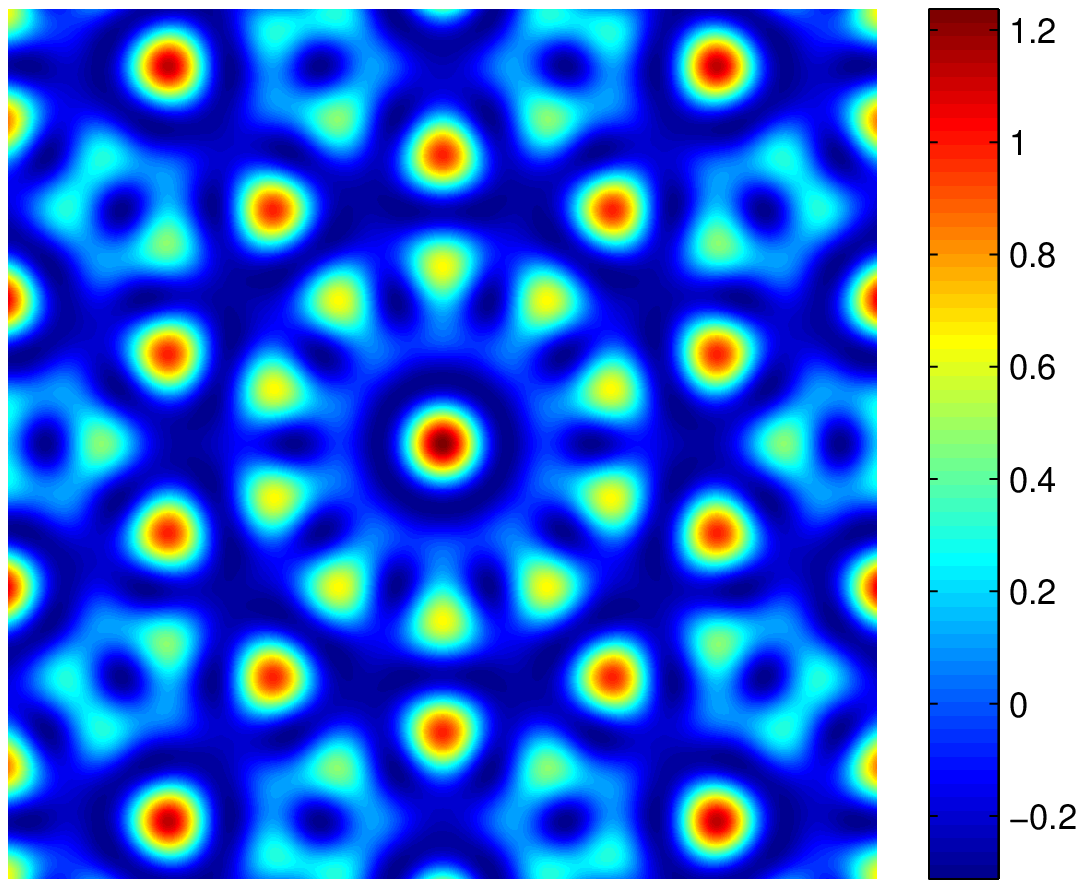}}
\subfigure[Effective density of B, $\Phi_B$]{\label{Subfigure: D phi_b}\includegraphics[scale = 0.32, bb = 100 280 500 580]{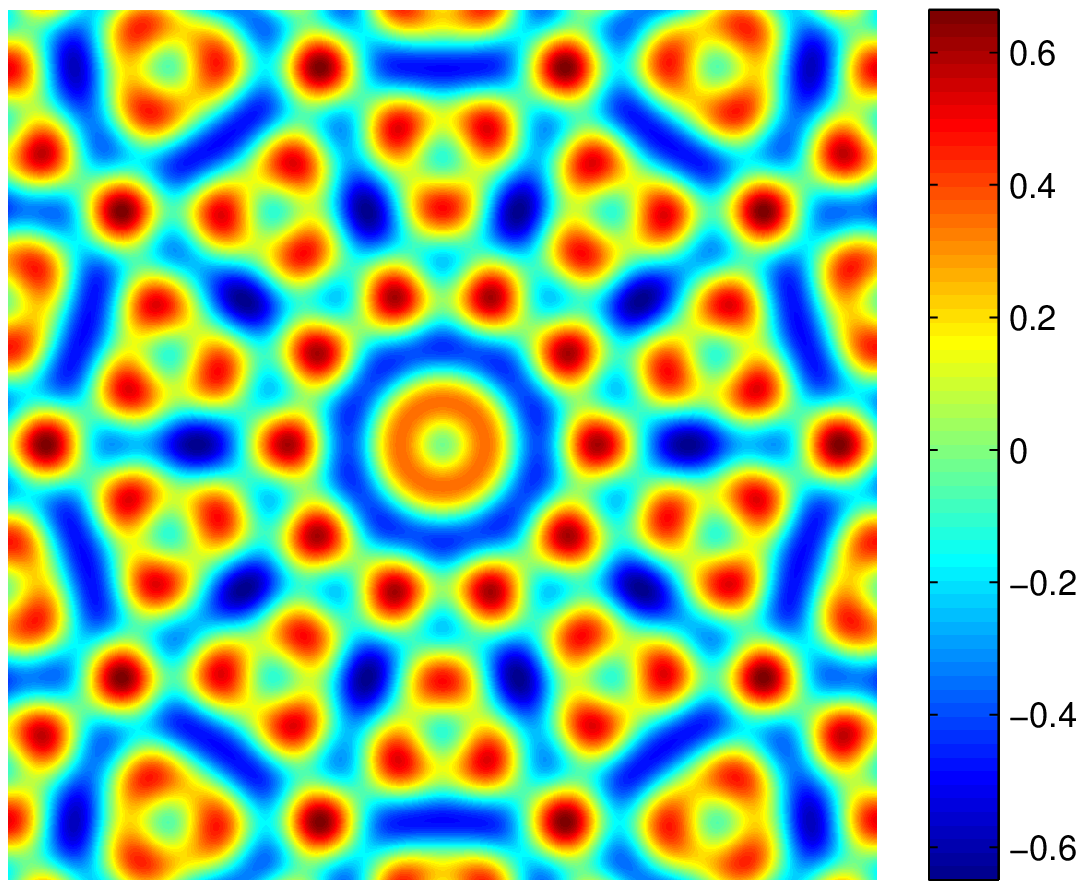}}
\subfigure[Effective density of C, $\Phi_C$]{\label{Subfigure: D phi_c}\includegraphics[scale = 0.32, bb = 100 280 500 580]{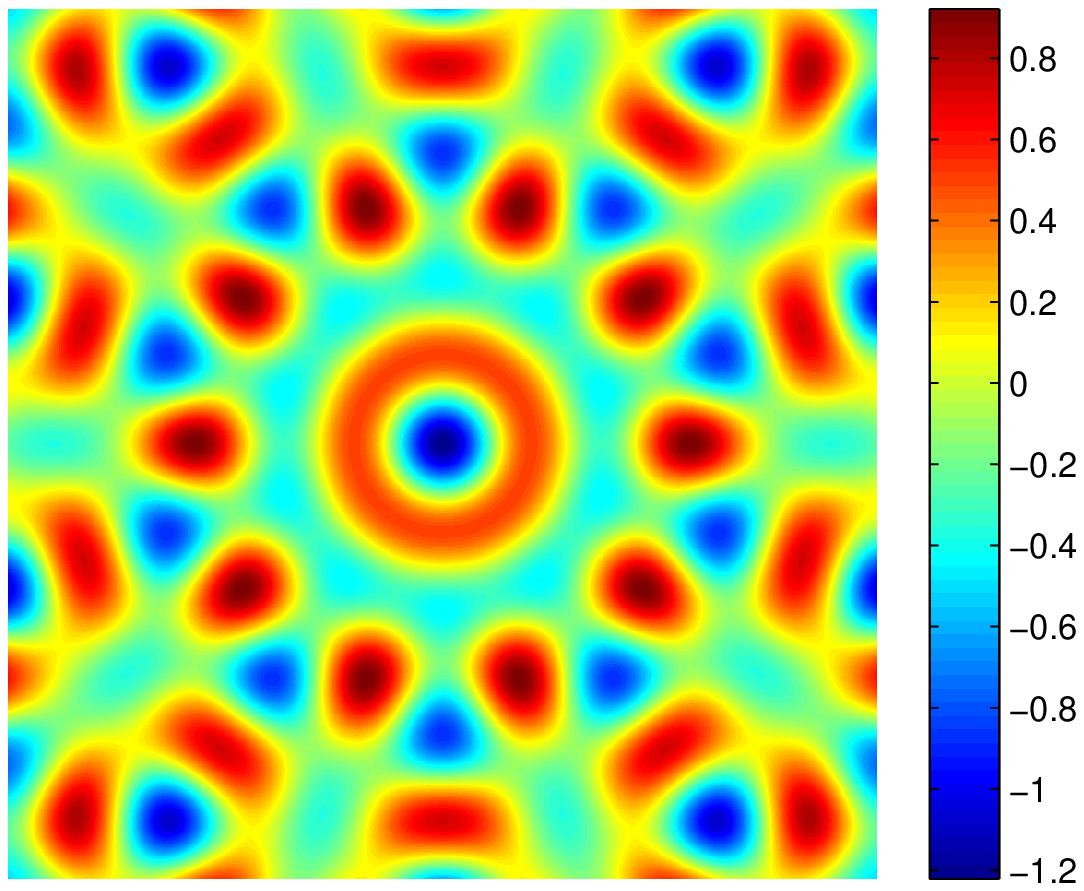}}
\caption{Illustration of the transformation from order parameters to real-space morphology in the case of decagonal quasicrystal obtained under $c= 80$, $q=2\cos(\pi/5)$, $t_0=0$, $g_0=0.8$, $g_1 = 2.2$, $g_2=0.2$, $t = 0$ and $\tau = 0$. FIG.~\ref{Subfigure: D psi} and FIG.~\ref{Subfigure: D phi} show the order parameters obtained from the direct minimization of the free energy functional \eqref{Equation: modified_model} using the numerical method proposed in Section~\ref{Subsection:
The strategy of optimizing the computational domain}. FIG.~\ref{Subfigure: D phi_a} to FIG.~\ref{Subfigure: D phi_c} show the effective densities recovered through \eqref{Equation: transformation from order parameter fields to effective density fields}. Then FIG.~\ref{Subfigure: D D real-morphology}, which is simply a copy of FIG.~\ref{Subfigure: D D} to show the real-space morphology of the decagonal quasicrystal, is obtained by finding out the dominant regions of the three components. The blue, green and red regions represent dominant regions of the component A, B and C respectively.}
\label{Figure: D show transform from order parameters to effective densities}
\end{figure}
\begin{figure}[!htbp]
\centering
\subfigure[\emph{DD} phase]{\label{Subfigure: DD DD real-morphology}\includegraphics[scale = 0.32, bb = 100 280 500 580]{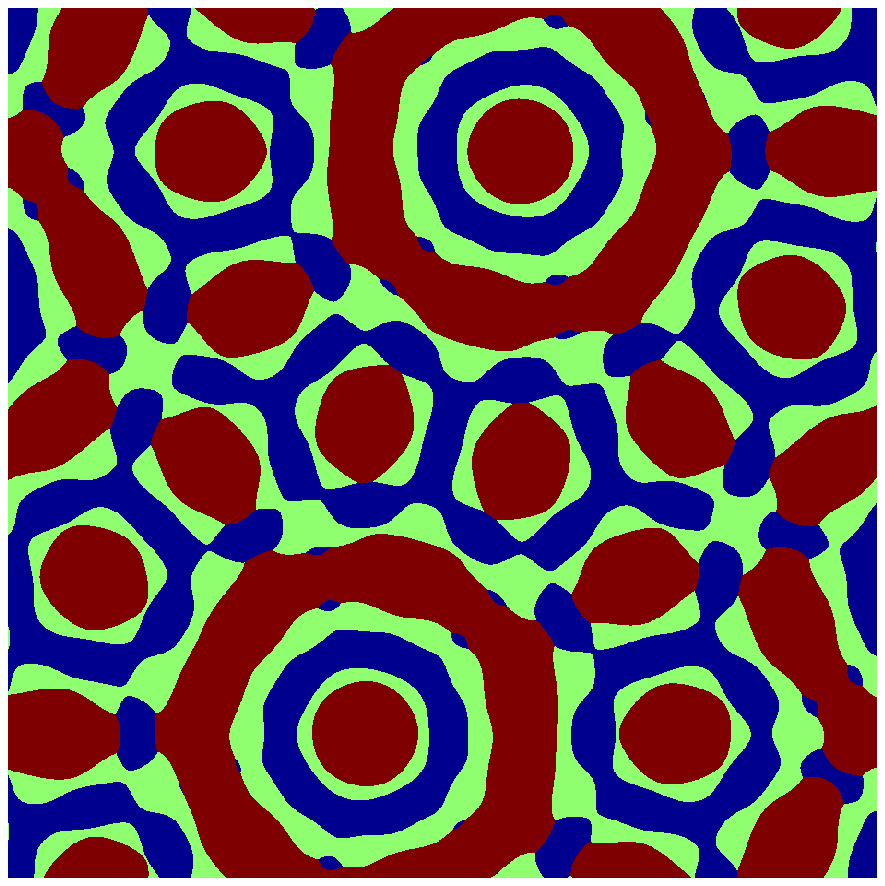}}
\subfigure[Order parameter $\psi$]{\label{Subfigure: DD psi}\includegraphics[scale = 0.32, bb = 100 280 500 580]{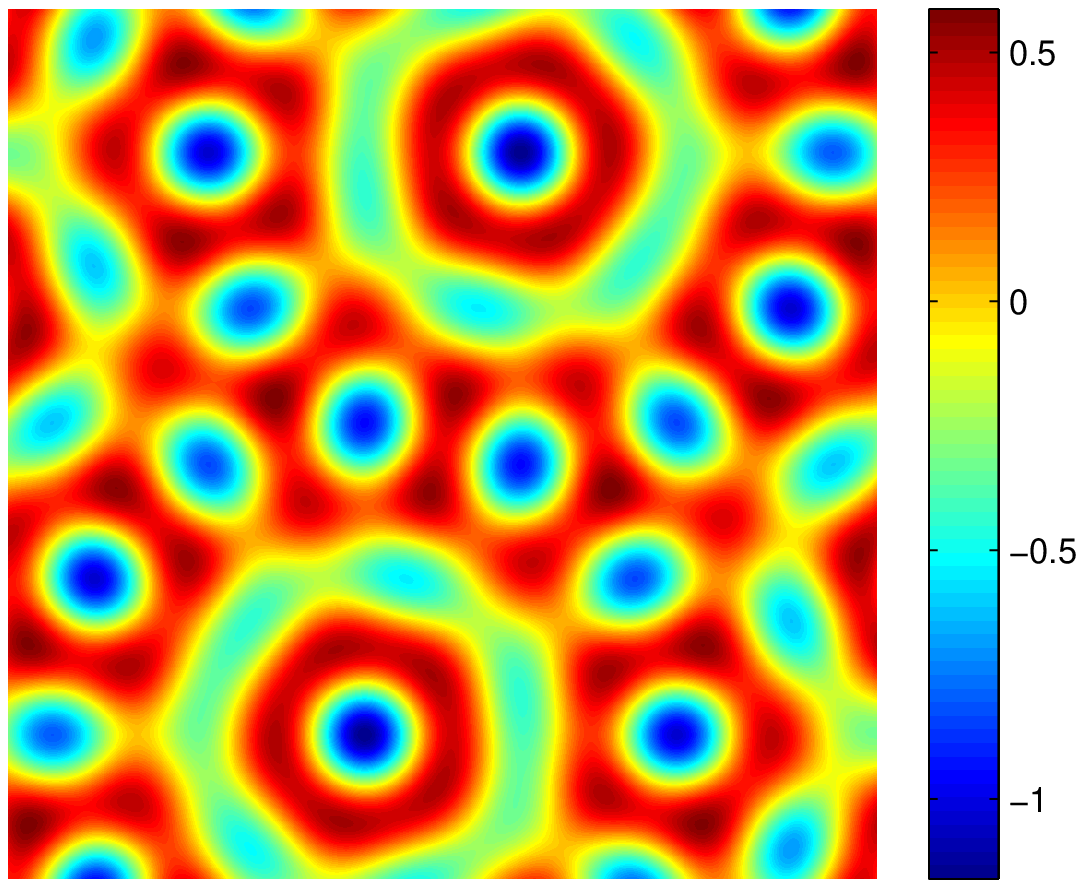}}
\subfigure[Order parameter $\phi$]{\label{Subfigure: DD phi}\includegraphics[scale = 0.32, bb = 100 280 500 580]{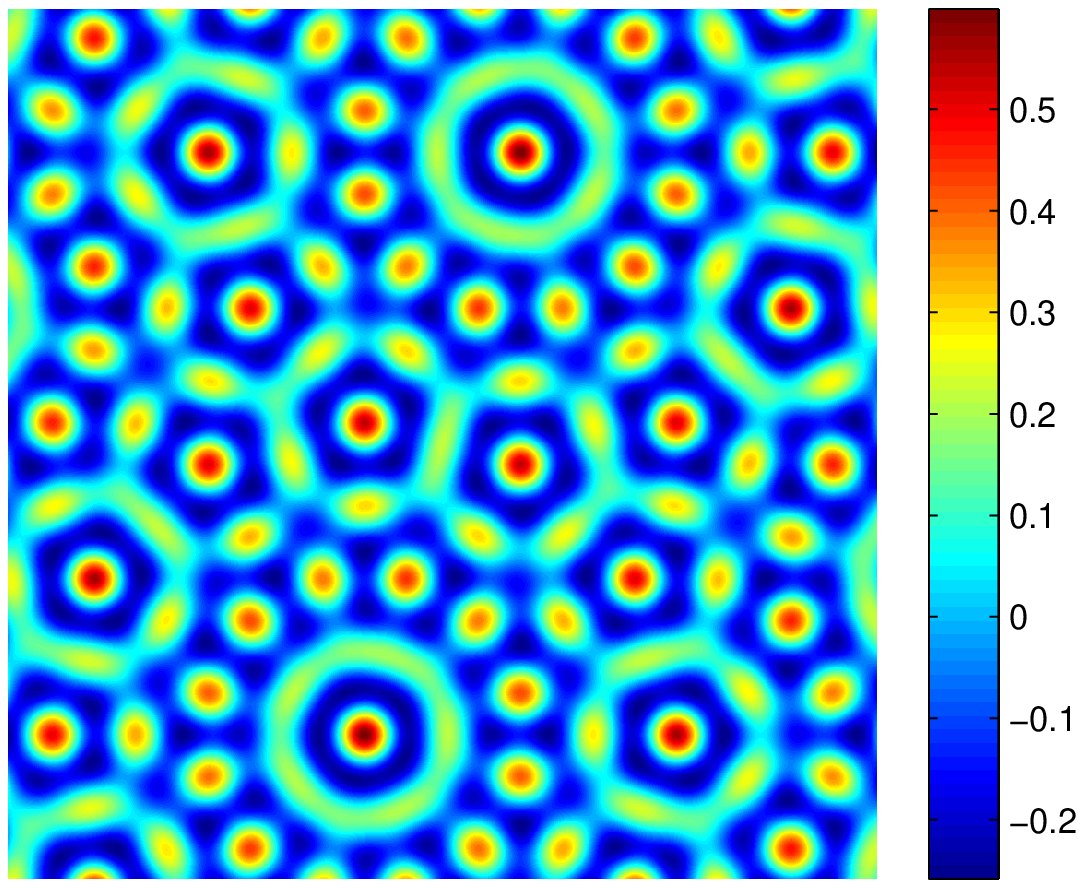}}\\
\subfigure[Effective density of A, $\Phi_A$]{\label{Subfigure: DD phi_a}\includegraphics[scale = 0.32, bb = 100 280 500 580]{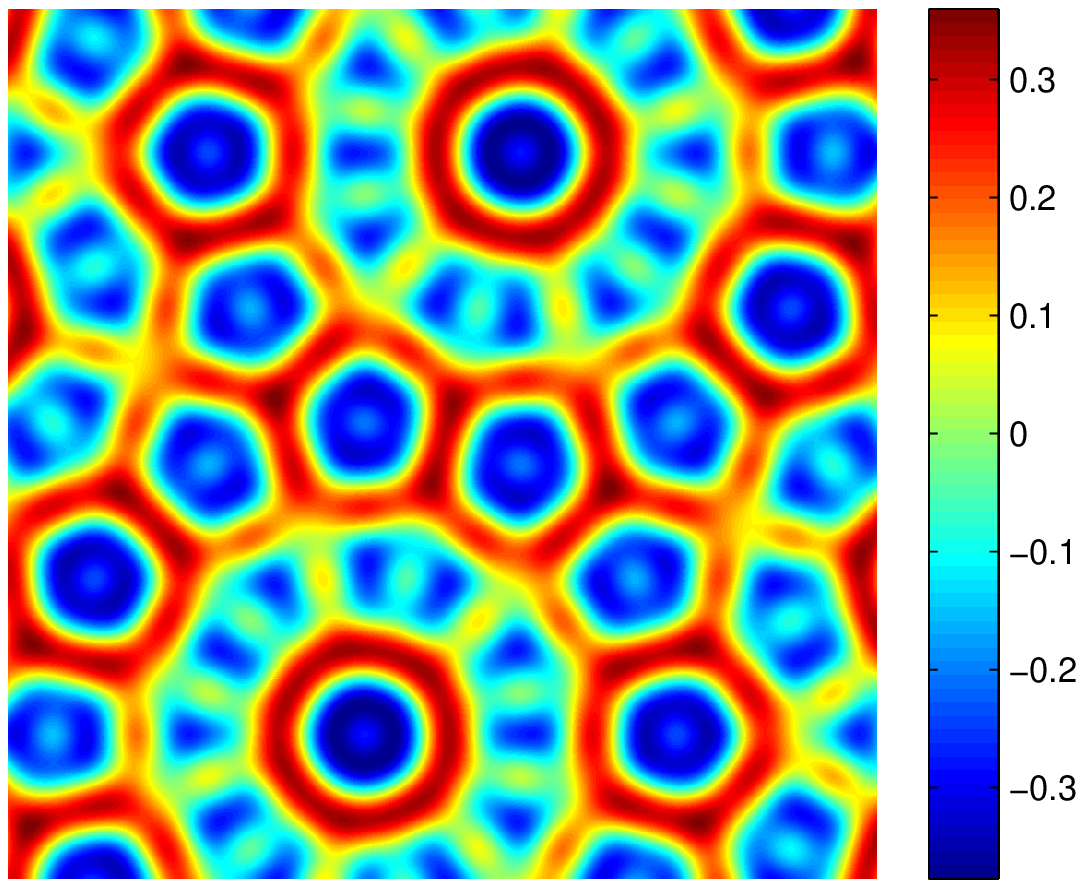}}
\subfigure[Effective density of B, $\Phi_B$]{\label{Subfigure: DD phi_b}\includegraphics[scale = 0.32, bb = 100 280 500 580]{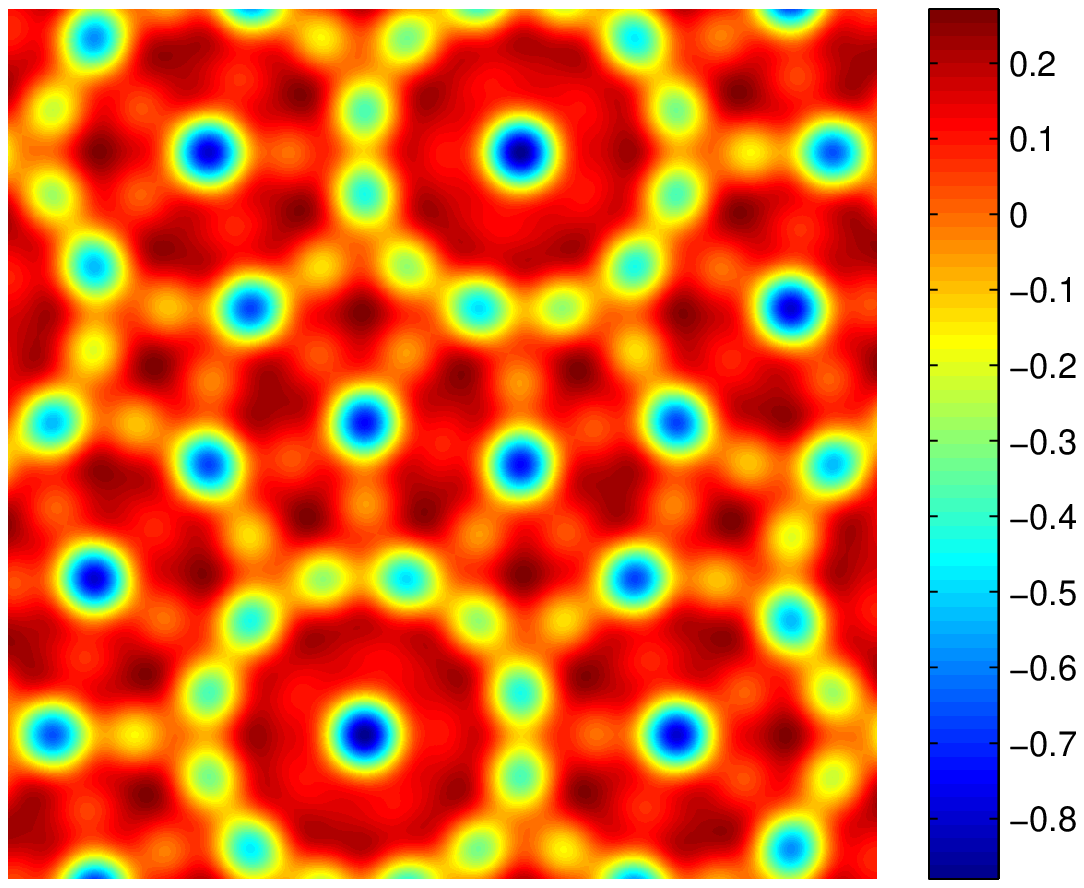}}
\subfigure[Effective density of C, $\Phi_C$]{\label{Subfigure: DD phi_c}\includegraphics[scale = 0.32, bb = 100 280 500 580]{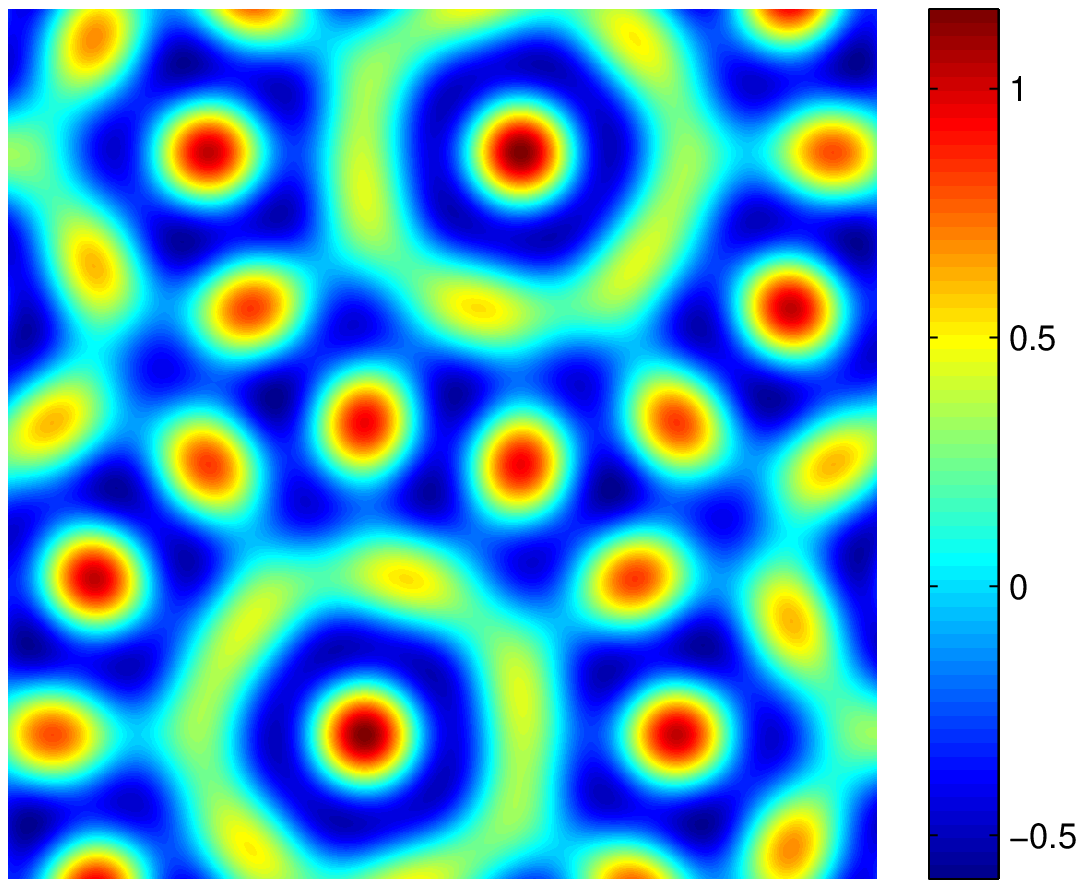}}
\caption{Illustration of the transformation from order parameters to real-space morphology in the case of dodecagonal quasicrystal obtained under $c= 80$, $q=2\cos(\pi/12)$, $q = 2\cos\frac{\pi}{12}$, $t_0=0$, $g_0=0.8$, $g_1 = 2.2$, $g_2=0.2$, $t = 0.3$ and $\tau = -0.2$. FIG.~\ref{Subfigure: DD psi} and FIG.~\ref{Subfigure: DD phi} show the order parameters obtained from the direct minimization of the free energy functional \eqref{Equation: modified_model} using the numerical method proposed in Section~\ref{Subsection:
The strategy of optimizing the computational domain}. FIG.~\ref{Subfigure: DD phi_a} to FIG.~\ref{Subfigure: DD phi_c} show the effective densities recovered through \eqref{Equation: transformation from order parameter fields to effective density fields}. Then FIG.~\ref{Subfigure: DD DD real-morphology}, which is simply a copy of FIG.~\ref{Subfigure: DD DD} to show the real-space morphology of the dodecagonal quasicrystal, is obtained by finding out the dominant regions of the three components. The blue, green and red regions represent dominant regions of the component A, B and C respectively.}
\label{Figure: DD show transform from order parameters to effective densities}
\end{figure}
\begin{figure}[!htbp]
\centering
\subfigure[\emph{D} phase at $t = 0,\tau =
0$]{\label{Subfigure: D D}\includegraphics[scale = 0.32, bb = 100 280 500 580]{phase_diagrams_D_80_D_t_0_tau_0.eps}}
\subfigure[\emph{L}$_3$ phase at $t = 1,\tau =
-0.5$]{\label{Subfigure: D L_3}\includegraphics[scale = 0.32, bb = 100 280 500 580]{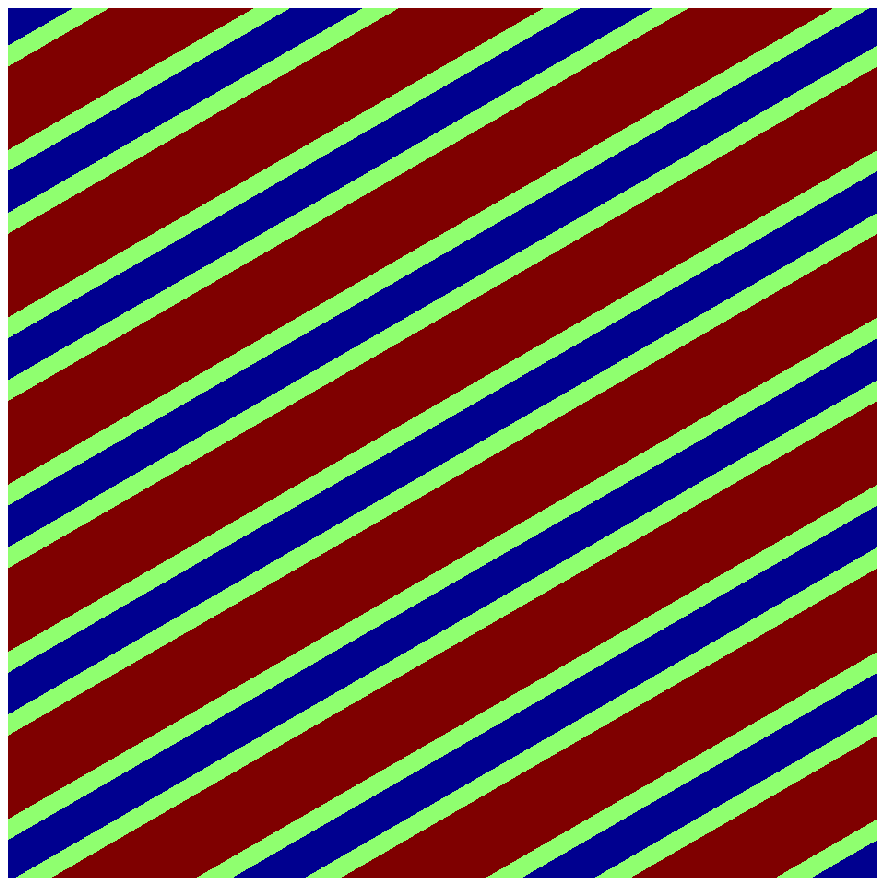}}
\subfigure[\emph{L}$_2$ phase at $t = -0.5,\tau =
1$]{\label{Subfigure: D L_2}\includegraphics[scale = 0.32, bb = 100 280 500 580]{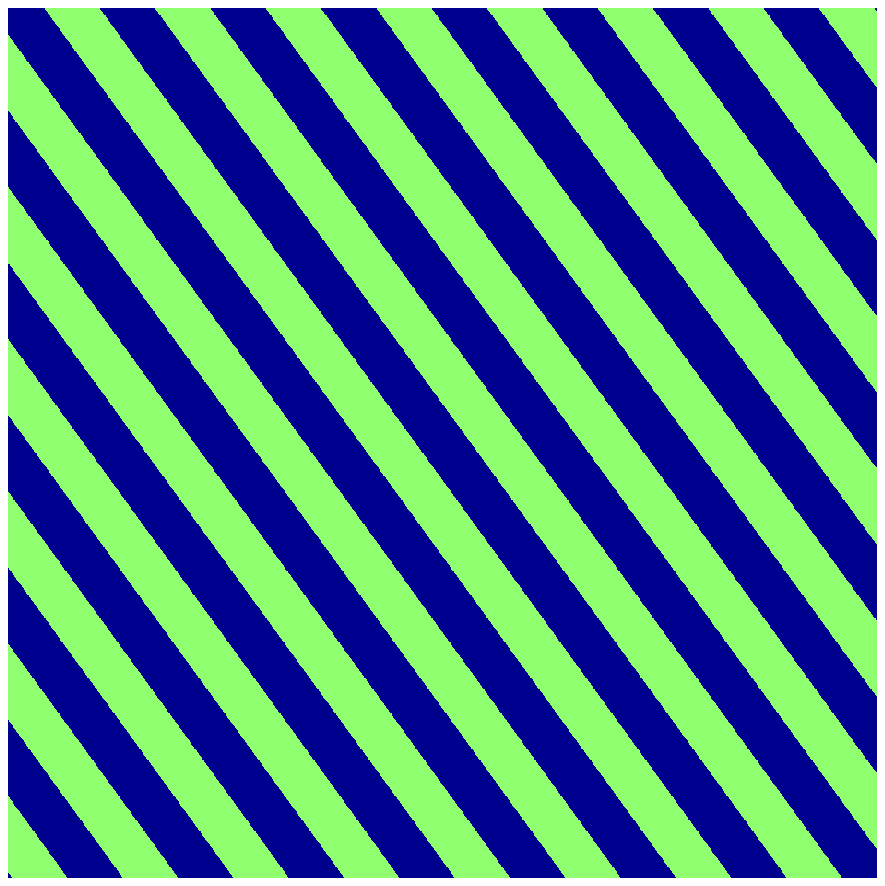}}
\subfigure[\emph{L+BI} phase at $t = -1.4,\tau =
1$]{\includegraphics[scale = 0.32, bb = 100 280 500 580]{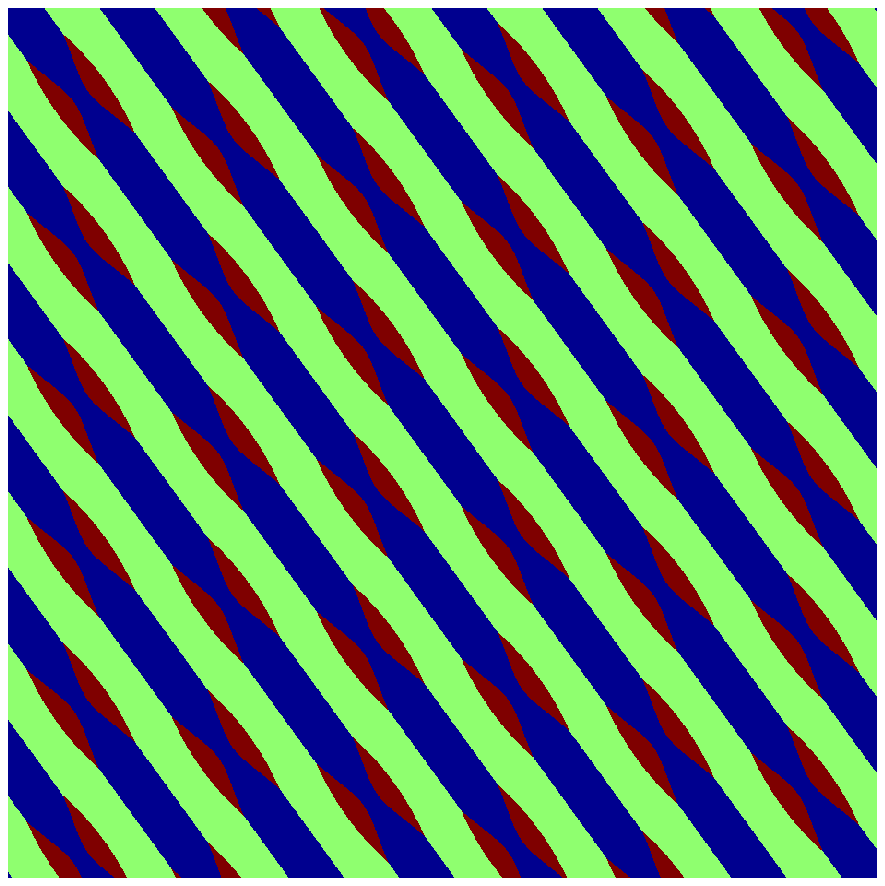}}
\subfigure[\emph{L+B} phase at $t = 0,\tau =
-2$]{\label{Subfigure: D L+B 1}\includegraphics[scale = 0.32, bb = 100 280 500 580]{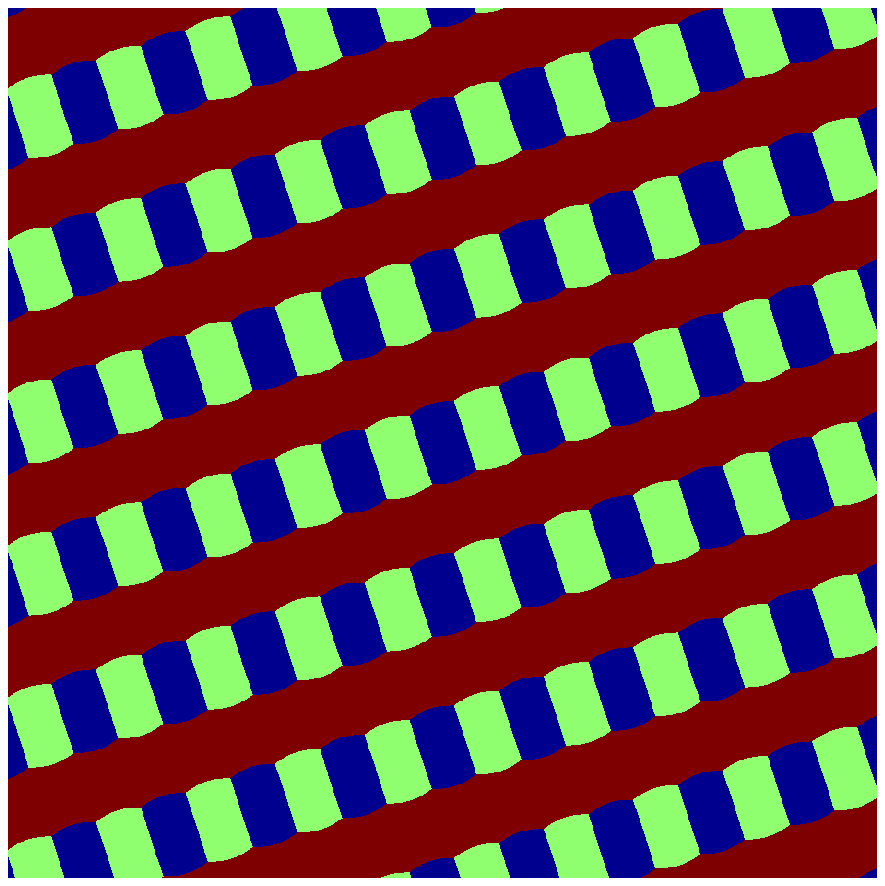}}
\subfigure[\emph{L+B} phase at $t = -0.1,\tau =
0.7$]{\label{Subfigure: D L+B 2}\includegraphics[scale = 0.32, bb = 100 280 500 580]{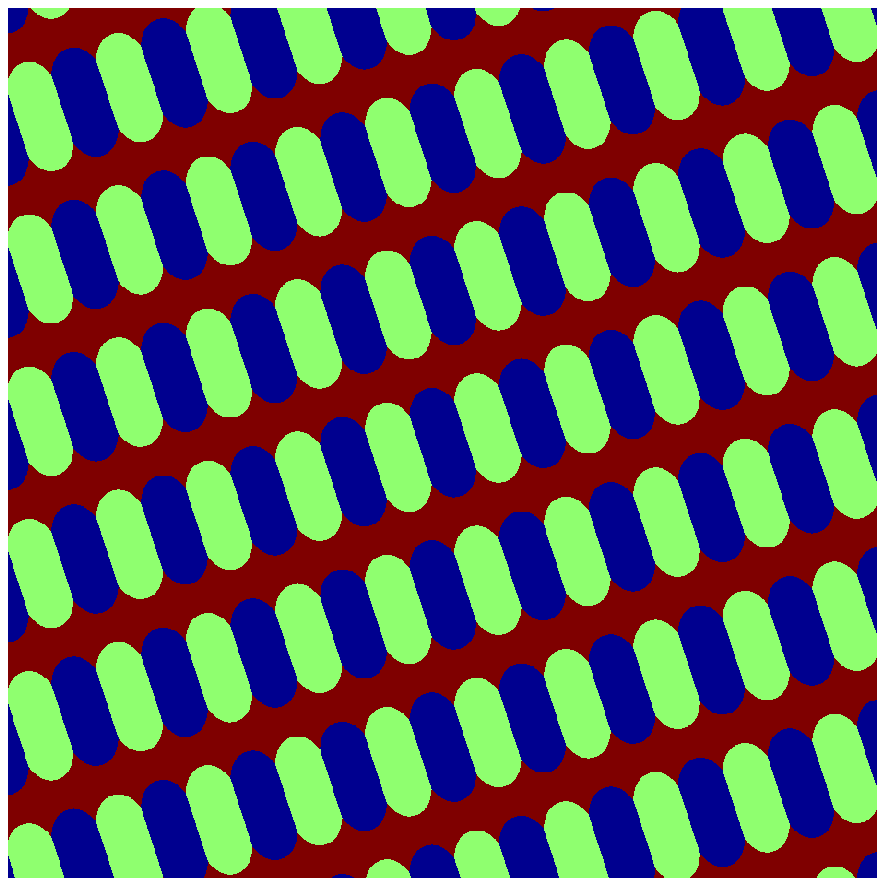}}
\subfigure[\emph{L+B} phase at $t = -2,\tau =
0$]{\label{Subfigure: D L+B 3}\includegraphics[scale = 0.32, bb = 100 280 500 580]{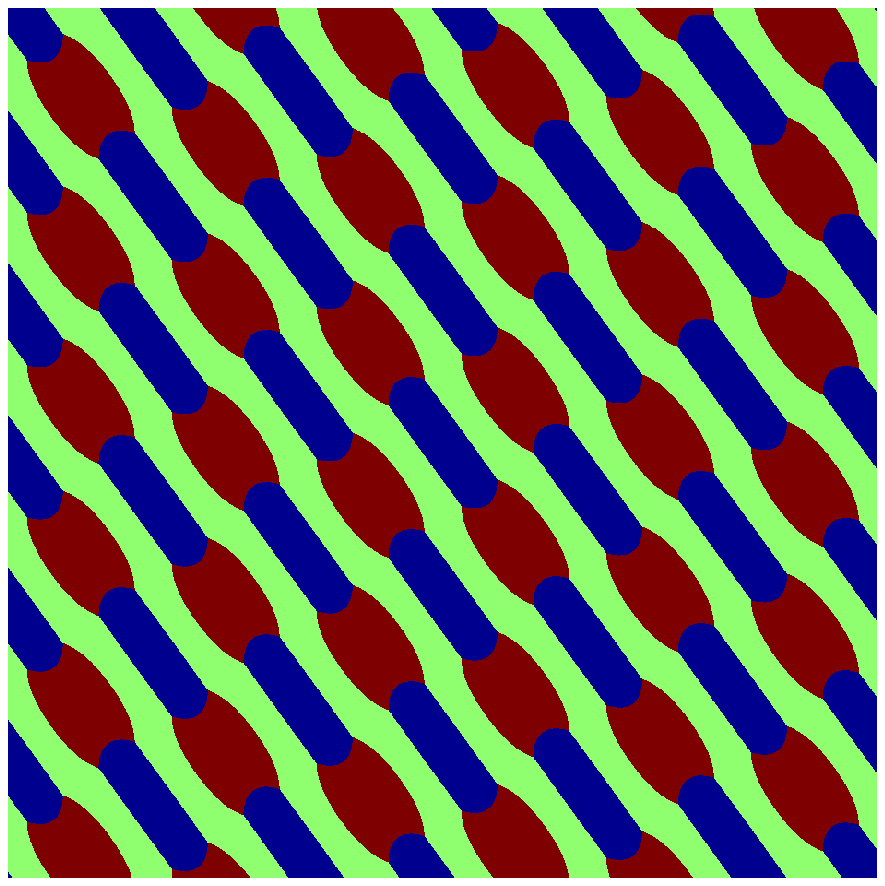}}
\subfigure[\emph{L+B} phase at $t = 0.7,\tau =
-0.1$]{\label{Subfigure: D L+B 4}\includegraphics[scale = 0.32, bb = 100 280 500 580]{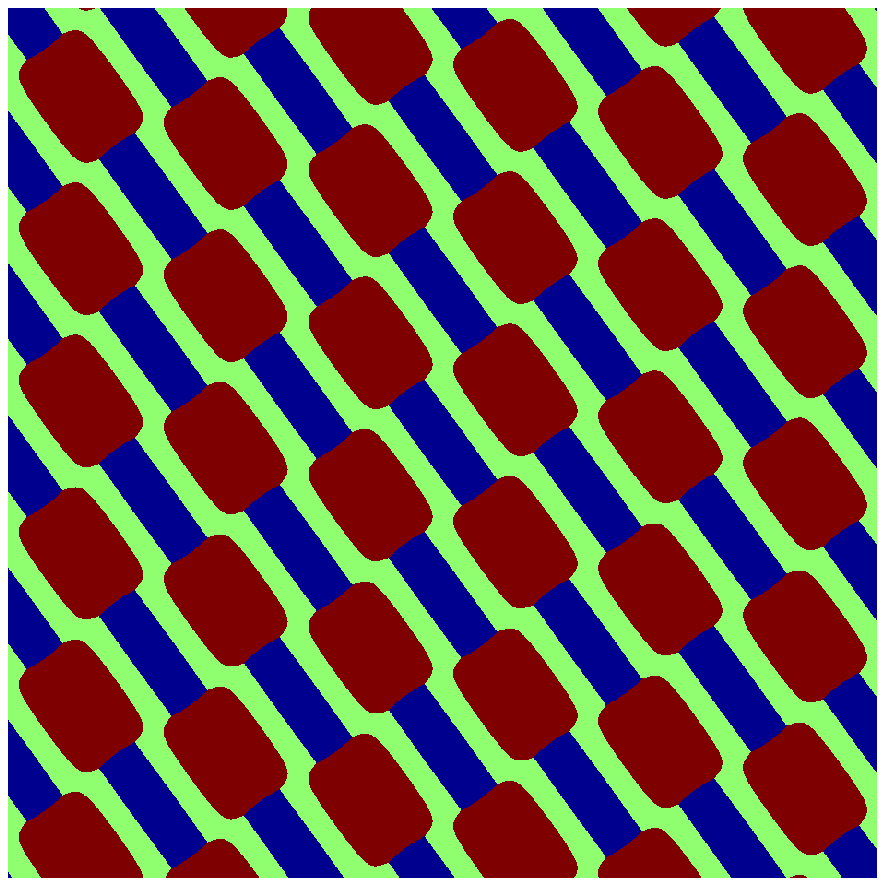}}
\caption{All phases in the phase diagram (see
FIG.~\ref{Subfigure: D_phase_diag c 80}) in the
decagonal case ($q=2\cos(\pi/5)$). The morphology is determined by plotting dominant regions of the
three components (see FIG.~\ref{Figure: D show transform from order parameters to effective densities} and FIG.~\ref{Figure: DD show transform from order parameters to effective densities} for illustrations).
The blue, green and red regions represent dominant regions of the component A, B and C respectively.
Note that \emph{L+B} phases have four different real-space morphologies in different regions of the phase diagram.}
\label{Figure: decagonal phases}
\end{figure}
\begin{figure}[!htbp]
\centering
\subfigure[\emph{DD} phase at $t = 0.3,\tau =
-0.2$]{\label{Subfigure: DD DD}\includegraphics[scale = 0.35, bb = 100 280 500 580]{phase_diagrams_DD_80_DD_t_3e-1_tau_-2e-1.eps}}
\subfigure[\emph{DD-Hmd} phase at $t = 0.46,\tau =
-0.5$]{\label{Subfigure: DDHmd}\includegraphics[scale = 0.35, bb = 100 280 500 580]{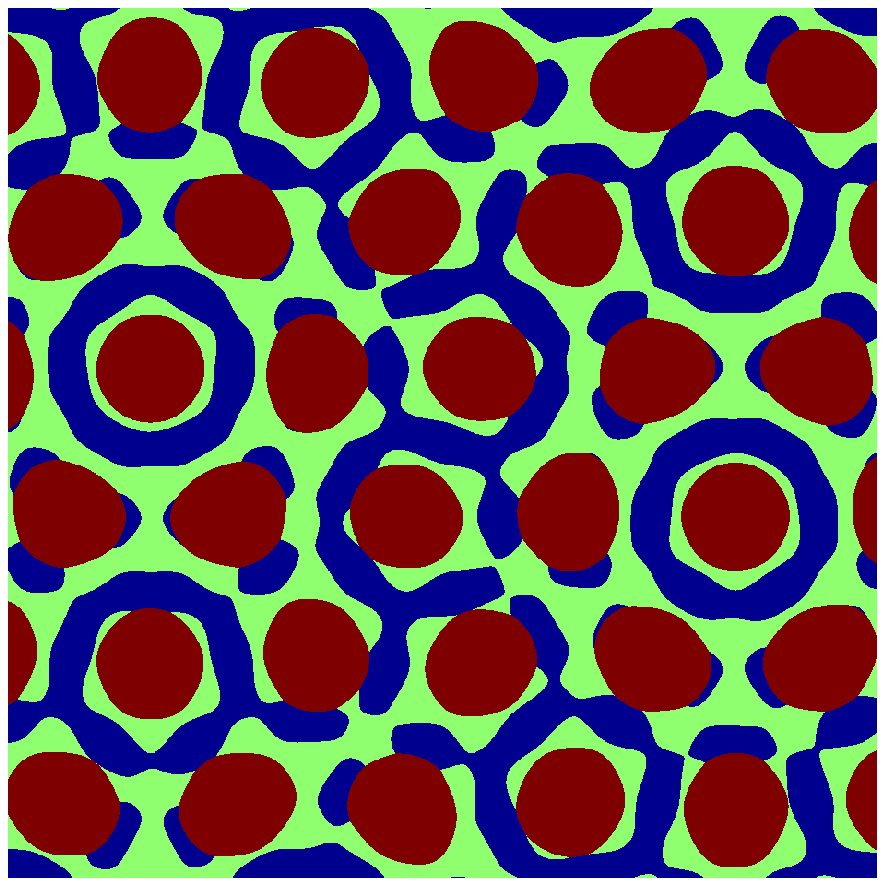}}
\subfigure[\emph{Hex+B} phase at $t = 0.5,\tau =
-0.3$]{\label{Subfigure: HexB}\includegraphics[scale = 0.35, bb = 100 280 500 580]{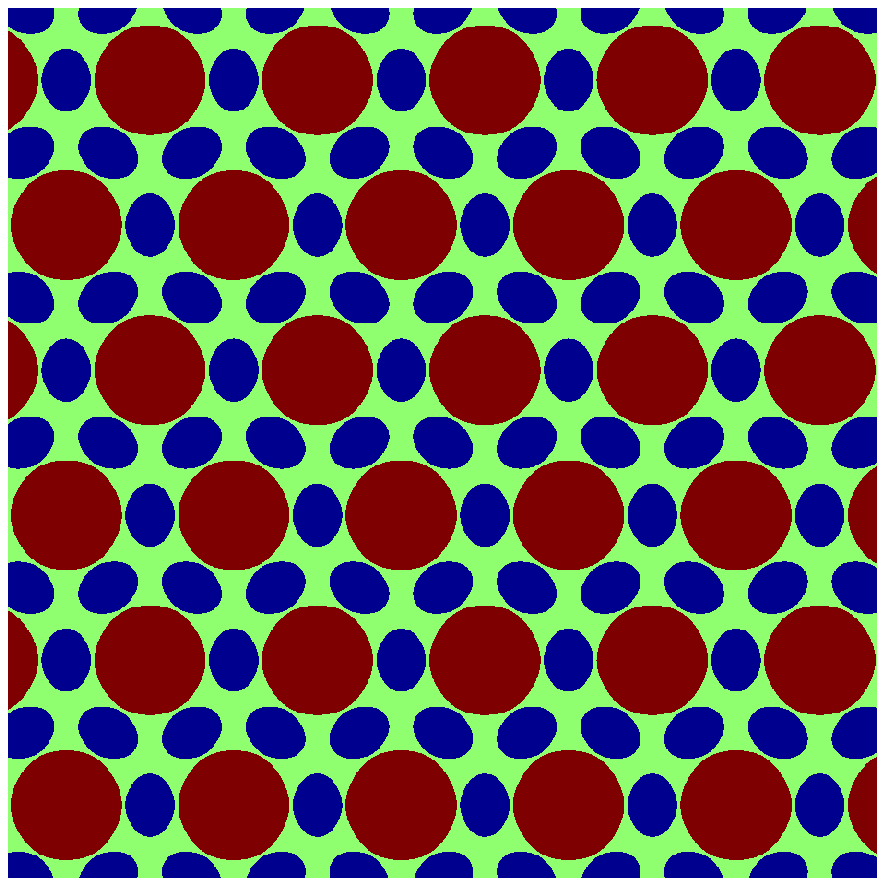}}
\subfigure[\emph{L+B} phase at $t = -0.1,\tau =
0$]{\includegraphics[scale = 0.35, bb = 100 280 500 580]{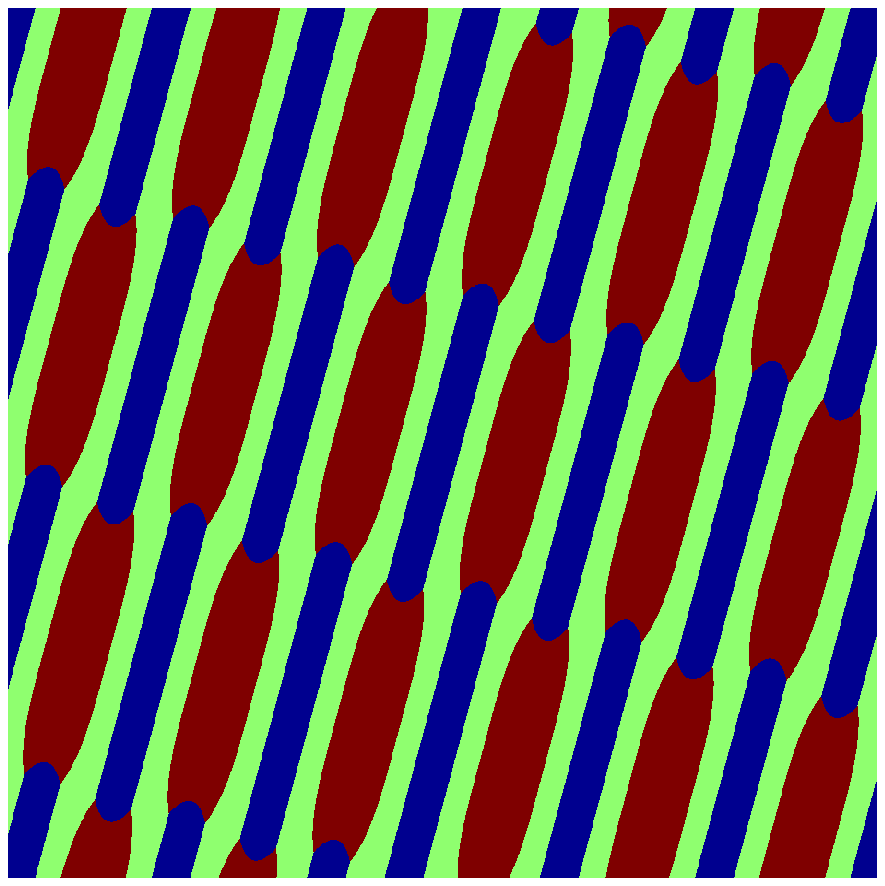}}
\caption{Selected phases in the phase diagram (see
FIG.~\ref{Subfigure: DD_phase_diag c 80}) in
the dodecagonal case ($q=2\cos(\pi/12)$). The morphology is determined by plotting dominant regions of the
three components (see FIG.~\ref{Figure: D show transform from order parameters to effective densities} and FIG.~\ref{Figure: DD show transform from order parameters to effective densities} for illustrations).
The blue, green and red regions represent dominant regions of the component A, B and C respectively.
For conciseness, the pictures for $L_3$ and $L_2$ phases are omitted.
See FIG.~\ref{Subfigure: D L_3} and FIG.~\ref{Subfigure: D L_2} in the decagonal case for reference.}
\label{Figure: dodecagonal phases}
\end{figure}

With the numerical method proposed in Section \ref{Subsection:
The strategy of optimizing the computational domain},
the phase diagrams are obtained in both decagonal and dodecagonal case
in the $t-\tau$ plane with $c = 80$, as is shown in FIG.~\ref{Figure: D DD phase diag c 80}.
Again, we note that only one $q$ is used throughout each individual phase diagram in computing all the ordered phases.
Morphology of the microphases in the both cases are exhibited in the FIG.~\ref{Figure: decagonal phases} and FIG.~\ref{Figure: dodecagonal phases} respectively.
As before, they are determined by plotting the dominant regions of components A, B and C \cite{tang2004morphology}.
As illustrations, we show in FIG.~\ref{Figure: D show transform from order parameters to effective densities} and FIG.~\ref{Figure: DD show transform from order parameters to effective densities} how we transform the fields of $\psi$ and $\phi$ into the effective densities $\Phi_A$, $\Phi_B$ and $\Phi_C$ and then into the real-space morphology in the cases of decagonal and dodecagonal quasicrystals respectively. Readers are referred to \eqref{Equation: transformation from order parameter fields to effective density fields} and the discussions associated to that in Section~\ref{Subsection: t-tau phase diagrams by the asymptotic study} for more details of performing this transformation.

In the decagonal case ($q=2 \cos(\pi/5)$), the new phase diagram is similar with the one in the limiting case shown in FIG.~\ref{Subfigure: D_phase_diag}.
The stability region for the decagonal quasicrystal \emph{D} slightly swells, while \emph{Hex} disappears in the phase diagram.
Moreover, a new transition phase named lamellae phase with beads at the interface, \emph{L+BI} for short, is discovered between \emph{L+B} and \emph{L}$_2$ phases with $\tau >0$.
We also note that the \emph{L+B} phase has four different shapes
in different regions of the phase diagram, as
FIGs.~\ref{Subfigure: D L+B 1}-\ref{Subfigure: D L+B 4} show.

The new phase diagram in the dodecagonal case ($q=\cos(\pi/12)$), i.e. FIG.~\ref{Subfigure: DD_phase_diag c 80}, significantly differs from the one in FIG.~\ref{Subfigure: DD_phase_diag}.
The stability region of \emph{DD} shrinks while three new phases are observed.
A new three-color lamellae phase \emph{L}$_3$ is favored when $\tau$ is sufficiently negative.
The hexagonal phase with beads, denoted by \emph{Hex+B}, appears in the place of \emph{Hex} in FIG.~\ref{Subfigure: DD_phase_diag}.
We note that $\phi = 0$ in \emph{Hex} when $c\rightarrow \infty$, while $\phi\not = 0$ in \emph{Hex+B} for finite $c$.
The morphology of \emph{Hex+B} is shown in FIG.~\ref{Subfigure: HexB}.
The third new phase, named dodecagonal quasicrystalline phase with hexagonal modulation, or \emph{DD-Hmd} for short, is not a periodic order.
FIG.~\ref{Subfigure: DDHmd} shows its real-space morphology.
It can be well described by a superposition of a perfect quasiperiodic order with dodecagonal symmetry and a periodic order with hexagonal symmetry.
It arises between the stability regions of the perfect dodecagonal quasicrystal phase (\emph{DD}) and the hexagonal phase with beads (\emph{Hex+B}), which can be viewed as a transition phase between the two.
To the best of our knowledge, it is the first time that such a modulated phase is reported in the numerical simulation.

The phase diagram in the dodecagonal case for $c = 80$ clearly shows the effect of $c$ on the phases behavior.
In the case of finite $c$, Fourier wave numbers other than $1$ (resp.~$q$) are allowed in the spectrum of $\psi$ (resp.~$\phi$).
As a result, in \emph{Hex+B} phase, $\phi$ is allowed to be non-zero.
Also, in the newly found \emph{L}$_3$ phase at the bottom of the phase diagram FIG.~\ref{Subfigure: DD_phase_diag c 80}, the magnitudes of principle wave vectors in the spectrum of $\psi$ (resp.~$\phi$) are not $1$ (resp.~$q$), which would be excluded in the limiting case.
We also remark that if no optimization is performed in the basis vectors, this \emph{L}$_3$ phase can hardly be captured.

\subsection{Phase diagram in $q$}
\begin{figure}[!htbp]
\centering
\includegraphics[scale = 0.5, bb = 300 180 600 300]{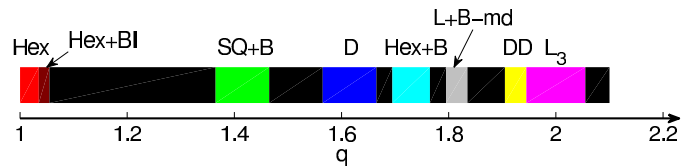}
\caption{One-dimensional phase diagram as a function of $q$ under $c = 80$, $t_0=0$, $g_0 = 0.2$, $g_1 = g_2 = 2.2$, $t = 0.3$ and $\tau = -0.2$. Phases with distinct real-space morphologies are marked with labels and different colors. The \emph{L+B} phase is favored in all the ranges colored as black; its label is omitted for conciseness. The stability ranges of
all the phases discovered here are also summarized Table~\ref{Table: stability range of phases in the q phase diagram}.}
\label{Figure: q 1d phase diagram}
\end{figure}

\begin{figure}[!htbp]
\centering
\subfigure[\emph{Hex} phase at $q=1.00$]{\includegraphics[scale = 0.35, bb = 100 280 500 580]{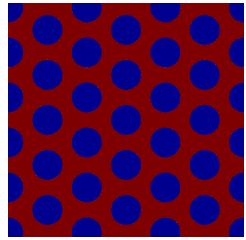}}
\subfigure[\emph{Hex+BI} phase at $q=1.05$]{{\label{Subfigure:
Hex+BI}}\includegraphics[scale = 0.35, bb = 100 280 500 580]{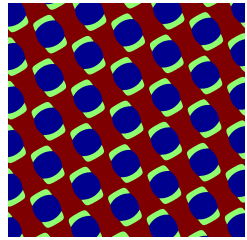}}
\\
\subfigure[\emph{SQ+B} phase at $q=1.40$]{{\label{Subfigure:
SQ+B}}\includegraphics[scale = 0.35, bb = 100 280 500 580]{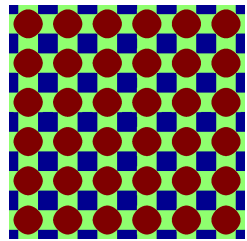}}
\subfigure[\emph{L+B-md} phase at $q=1.80$]{{\label{Subfigure:
L+B-md}}\includegraphics[scale = 0.35, bb = 100 280 500 580]{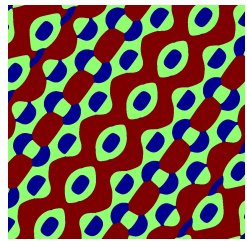}}
\caption{Selected phases in the phase diagram in
terms of $q$ for fixed $c = 80$, $t_0=0$, $g_0 = 0.2$, $g_1 = g_2 = 2.2$, $t = 0.3$ and $\tau = -0.2$. The morphology is determined by plotting dominant regions of the
three components (see FIG.~\ref{Figure: D show transform from order parameters to effective densities} and FIG.~\ref{Figure: DD show transform from order parameters to effective densities} for illustrations).
The blue, green and red regions represent dominant regions of the component A, B and C respectively. The stability ranges of
all the phases discovered in $q\in[1.00,2.10]$ are summarized in FIG.~\ref{Figure: q 1d phase diagram} and Table~\ref{Table: stability range of phases in the q phase diagram}.}
\label{Figure: q phases}
\end{figure}

In the above discussion, we have noted that $q$ is a system-specific parameter that determines the ratio of two characteristic length scales between $\psi$ and $\phi$.
It is interesting to study how the parameter $q$ may influence the
pattern formation.
In the sequel, we fix $c = 80$, $t_0=0$, $g_0 = 0.2$, $g_1 =
g_2 = 2.2$, $t = 0.3$ and $\tau = -0.2$, and apply the numerical
method developed in Section \ref{Subsection: The strategy of optimizing the computational domain} to find out the ground states for $q\in[1.00,2.10]$ with resolution $0.01$.
The choices of initial configurations in the computations are much similar with those in FIG.~\ref{Figure: D_initial configurations} and FIG.~\ref{Figure: DD_initial configurations}.
The stability ranges of
ground states with distinct real-space morphology are shown in FIG.~\ref{Figure: q 1d phase diagram}.
Besides the patterns that have been already observed in FIG.~\ref{Figure: decagonal phases} and FIG.~\ref{Figure: dodecagonal phases}, the newly discovered structures are exhibited in FIG.~\ref{Figure: q phases}.
Periodic phases we have found before such as \emph{Hex} and $\emph{Hex+B}$, and quasicrystalline phases \emph{D} and \emph{DD} are observed once again, along with three new phases.
The hexagonal phase with beads at the interface, \emph{Hex+BI}
for short, occurs as a transition between \emph{Hex}
phase that is stable in $q\in[1.00,1.03]$ and \emph{L+B} at
$q\in[1.06,1.36]$. Its real-space morphology is shown in FIG.~\ref{Subfigure: Hex+BI}.
Square phase with beads (\emph{SQ+B}) is a periodic phase with four-fold symmetry found in $q\in[1.37,1.46]$.
See FIG.~\ref{Subfigure: SQ+B}.
A third phase (see FIG.~\ref{Subfigure: L+B-md}) observed in $q\in[1.80,1.83]$ is neither a periodic order, nor any quasicrystalline orders with rotational symmetry.
We claim that it is a superposition of \emph{L+B} phase and an incommensurate modulation, thus named as \emph{L+B-md}.
It is still not clear, however, whether such phase can be
observed in the experiments and whether it implies another type
of stable phase which we are not yet able to capture in the
current simulations (for example, three-dimensional ordered phases could be favored in this range).


\begin{table}[h]
\begin{center}
\caption{Critical length scale $q^*$ and stability ranges in $q$ of various phases in FIG.~\ref{Figure: q 1d phase diagram}}
\label{Table: stability range of phases in the q phase diagram}
\begin{tabular}{lll}\toprule[1pt]
Phases & Critical length scale $q^*$ &Stability ranges in $q$ phase diagram\\\toprule[0.5pt]
\emph{Hex}&$1$&$[1.00,1.03]$\\[1pt]
\emph{Hex+BI}&N/A&$[1.04,1.05]$\\[1pt]
\emph{SQ+B}&$2\cos(\pi/4)\approx 1.414$ (4-fold symmetry)  &$[1.37,1.46]$\\[1pt]
\emph{D}&$2\cos(\pi/5)\approx 1.618$  (see Ref.~\cite{steurer2009crystallography}) &$[1.57,1.66]$\\[1pt]
\emph{Hex+B}&$2\cos(\pi/6)\approx 1.732$ (6-fold symmetry)  &$[1.70,1.76]$\\[1pt]
\emph{L+B-md}&N/A&$[1.80,1.83]$\\[1pt]
\emph{DD}&$2\cos(\pi/12)\approx 1.932$ (see Ref.~\cite{steurer2009crystallography})  &$[1.91,1.94]$\\[1pt]
\emph{L}$_3$&$2$&$[1.95,2.05]$\\[1pt]
\emph{L+B}&N/A&$[1.06,1.36]$, $[1.47,1.56]$, $[1.67,1.69]$,\\[1pt]
&&$[1.77,1.79]$, $[1.84, 1.90]$, $[2.06,2.10]$\\[1pt]
\bottomrule[1pt]
\end{tabular}
N/A implies that the critical length scale $q^*$ can not be well defined in these cases.
\end{center}
\end{table}

To further explain the stability affected by the characteristic length scale,
we define the critical length scale $q^*$ of a phase to be the length scale geometrically favored by the desired (i.e.~standard) configuration of its wave vectors.
For example, in the decagonal and dodecagonal quasicrystalline
orders, the critical length scale $q^*$'s are $2\cos\frac{\pi}{5}$ and
$2\cos\frac{\pi}{12}$ respectively, see Section \ref{Subsection:
t-tau phase diagrams by the asymptotic study} or
Refs.~\cite{steurer2009crystallography}.
Table~\ref{Table: stability range of phases in the q phase
diagram} gives the critical length scale $q^*$'s and the stability
intervals of various patterns.
Note that the $q^*$ for \emph{L+B} phases is not well defined, since their Fourier wave vectors can be adjusted to fit a wide range of $q$.
It is observed that the stability ranges of most phases in
Table~\ref{Table: stability range of phases in the q phase
diagram} are neighborhoods of their critical length scale $q^*$'s.

\begin{figure}[!htbp]
\centering
\includegraphics[scale = 0.7]{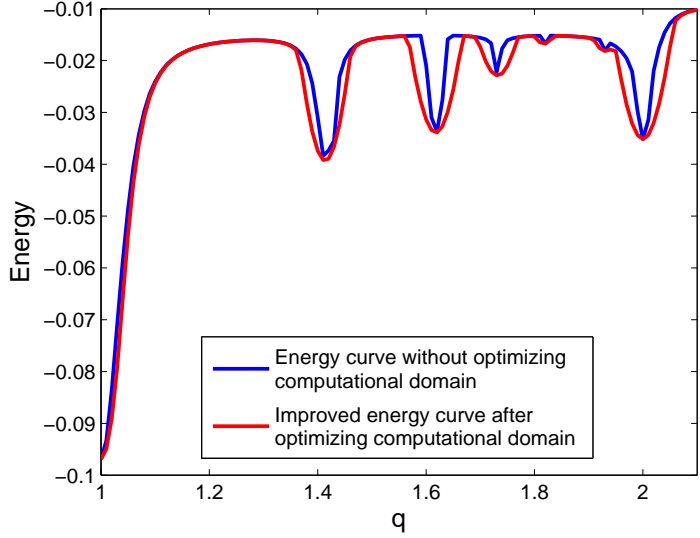}
\caption{Curves of the ground-state energy as functions of $q$
for $c = 80$, $t_0=0$, $g_0 = 0.2$, $g_1 = g_2 = 2.2$, $t = 0.3$
and $\tau = -0.2$. The blue curve represents the energy obtained
without optimizing computational cell, while the red one is the
improved energy curve after optimizing computational cell. One
can see the strategy of optimizing computational cell makes improvement in energy in the basins.}
\label{Figure: q energy}
\end{figure}

The ground-state energy as a function of $q$ is plotted as the
red curve in FIG.~\ref{Figure: q energy} where the shape and size
of the computational cell have been optimized by the variable cell method.
It is clear that the energy will be significantly lowered when $q$ is close to $q^*$'s of the phases in Table~\ref{Table: stability range of phases in the q phase diagram} whenever $q^*$ is well-defined, forming ``basins" on the energy curve.
As a comparison, we also plot the energy obtained without optimizing computational domain as the blue curve in the same figure.
We can see that the red curve is always below the blue one, and
it makes considerable improvement in evaluating energy in the basins compared to the blue curve.
This implies that the variable cell method can better capture phase behavior of the
model than a simple minimization.
The reason is that, when $q$ is slightly inconsistent with the
$q^*$ of a particular phase, the energy value of the phase can be lowered by slightly adjusting the basis vectors, especially their lengths.
When $q$ is far away from $q^*$, however, chances become small
such that the phase corresponding to $q^*$ can be distorted to fit $q$.

\section{Conclusion and Discussion\label{Section: Conclusion}}
In this article, we develop a coupled-mode Swift-Hohenberg model
with two order parameters to study the stability of soft quasicrystals in three-component systems.
Two characteristic length-scales are added into the original
Swift-Hohenberg free energy functional through the positive-definite
gradient terms acting as penalizing factors on Fourier wave numbers.
In order to investigate ordered structures, a computational framework is also proposed combining the projection method and the variable cell approach.
It enables accurate calculations of the free energy of ordered patterns, leading to
phase diagrams with more accurate phase boundaries.
Moreover, it shows the potential to discover more complex
phases, such as incommensurate modulation structures.
The generalized model as well as the numerical framework provide a powerful
tool for investigating the phase behavior of ordered patterns on the mean-field level.

Employing this tool, we study the phase behavior of the
coupled-mode model under the soft constraints on wave numbers (i.e.~$c=80$),
and compare it with that in the limiting case $c\rightarrow +\infty$ (i.e.~infinitely stiff constraints),
which is obtained by the two-mode approximation method.
The value of $q$'s are selected to form
the decagonal ($2\cos(\pi/5)$) and dodecagonal ($2\cos(\pi/12)$) symmetries, respectively.
Due to high computational cost, we only considered the two-dimensional
periodic structures and quasicrystals whose Fourier spectrum can be embedded into a periodic lattice in the four-dimensional space.
A number of two-dimensional ordered
structures emerge in our calculations, including
periodic crystals, 10-, 12-fold symmetric quasicrystals and
modulated structures.
By comparing the free energies of these candidates, phase
diagrams are obtained in the $t-\tau$ plane in both cases of
$c=80$ and $c\rightarrow +\infty$.
It is predicted that decagonal quasicrystal ($q=2\cos(\pi/5)$)
and dodecagonal quasicrystal ($q=2\cos(\pi/12)$) can
become ground states for both $c$'s.
The differences of the phase diagrams in these two cases are also emphasized
to show the important role of $c$ and the optimization of the computational domain.
In particular, we remark that the classic two-mode approximation analysis a priori assumes the order parameters favor specific length scales
$k=1$ and $k=q$, and is not sufficient to capture the true phase behavior of models of this type.

Furthermore, the role of the ratio of two characteristic length-scales, $q$,
on the pattern formation is also studied.
A one-dimensional phase diagram as a function of $q$ with
other system parameters fixed is obtained, as is shown in FIG.~\ref{Figure: q 1d phase diagram}.
From the phase diagram, we can see that the pattern formation
depends on the critical length scale $q^*$, which is
determined by the standard configurations of wave vectors of
the corresponding ordered patterns.
For most ordered patterns, their stable regions are simply neighborhoods of their corresponding critical length
scales. A modulated structure is discovered for $q\in[1.80, 1.83]$. It could be an indication of new phases in experiments or shows the limitation of the current study; for example, a three-dimensional structure might be favored over all the two-dimensional structures in this range, which we have not studied.
It should be emphasized that when varying the value of $q$, the
variable cell method plays an important role of accurately calculate the free energy.

These results provide a good understanding of the rich phase
behavior on the mean-field level in the coupled-mode Swift-Hohenberg model for
three-component systems.
It confirms that the existence of two
characteristic length scales and sufficiently strong three-body
interactions can account for the stability of soft quasicrystals
\cite{lifshitz2007soft}.
In the current work, the 10-, 12-fold symmetric quasicrystals, and some modulated structures can become stable phases.
To this date, in the most experiments of soft matters, quasicrystals have been
observed in dodecagonal symmetry, whereas decagonal quasicrystal
have not been reported. However, it is still worth expecting the
decagonal symmetric quasicrystals, and even modulated structures in the
multi-component soft-matter systems.

The numerical framework we developed, combining the projection method with the variable cell method,
is suitable for studying periodic and quasiperiodic orders in the current model.
It also works for a larger class of mean-field energy functionals.
For example, if the differential terms in (\ref{Equation:
modified_model}) are replaced by convolution-type interaction kernels \cite{barkan2011stability,rottler2012morphology}
\begin{equation}
\begin{split}
f[\psi,\phi]=\frac{1}{V}&\left[\int\mathrm{d}\mathbf{r}_1\int\mathrm{d}\mathbf{r}_2\,\psi(\mathbf{r}_1)C_1(\mathbf{r}_1-\mathbf{r}_2)\psi(\mathbf{r}_2)
+\phi(\mathbf{r}_1)C_2(\mathbf{r}_1-\mathbf{r}_2)\phi(\mathbf{r}_2)\right.\\
&\left.+\int\mathrm{d}\mathbf{r}\,\tau\psi^2+g_0\psi^3+\psi^4+t\phi^2+\phi^4-g_1\psi^2\phi-g_2\psi\phi^2\right],
\end{split}
\label{Equation: convolution-type_models}
\end{equation}
our method works as well.
Indeed, by Parseval's identity, the convolution can be rewritten as
\begin{equation*}
\frac{1}{V}\int\mathrm{d}\mathbf{r}_1\int\mathrm{d}\mathbf{r}_2\,\psi(\mathbf{r}_1)C_1(\mathbf{r}_1-\mathbf{r}_2)\psi(\mathbf{r}_2)
=
\frac{C}{V}\int\mathrm{d}\mathbf{k}\,\hat{C}_1(\mathbf{k})|\hat\psi(\mathbf{k})|^2,
\end{equation*}
where $C>0$ is a constant coming out from the Fourier transform.
It is essentially in the same form as (\ref{Equation: apply parseval's identity to the modified model}), which can be dealt with in exactly the same way.

A potential drawback of our numerical method is that, although it allows to perform minimization with larger degrees of freedom, it is not of high accuracy, especially when $c$ and $N$ are small.
It is well-known that the ordinary spectral method enjoys spectral
accuracy in dealing with periodic problems in general, but it
might be not the case in our quasiperiodic settings.
The part of energy, for which the wave vectors in (\ref{Equation: discretization point in projection method}) do not account, decreases slowly as $N$ grows.
In other words, there is still a considerable amount of energy contributed by the wave vectors with high indices, which are not necessarily of high frequencies.
In the current work, we set $c = 80$ and use $N = 16$ bases in the direction of each basis vector.
It is sufficient to obtain energy of the ground states with four-
to five-digit precision and thus can be used in constructing the phase diagrams.
However, when $c$ is smaller, we need larger $N$, which is not
always affordable for four- or higher-dimensional computations in the projection method.
Recall that the computational cost in one step of steepest descent iteration grows as $O(N^4\log N)$ in computing decagonal and dodecagonal quasicrystals.
The problem of low accuracy can be model-specific, since it does not occur in our earlier work on the Lifshitz-Petrich model \cite{jiang2014LPmodel}.
In fact, the penalizing effect of the differential terms in the Lifshitz-Petrich model is much stronger than that in (\ref{Equation: modified_model}).
The former one is proportional to $|\mathbf{k}|^4$ as $|\mathbf{k}|\rightarrow+\infty$, while the latter one grows as $|\mathbf{k}|^8$.
This implies that the Fourier wave numbers far away from
the prescribed length scales (i.e., $1$ and $q$) are suppressed more
strongly in the Lifshitz-Petrich model than in the current model.
As a result, their contribution to the energy becomes less significant in the Lifshitz-Petrich model.
Nevertheless, it is still unknown how to achieve higher accuracy in the quasiperiodic setting in the current setting.

\section*{Acknowledgement}
The work is supported by the NSFC projects 11421110001, 11421101, 21274005, 11401504, and 91430213.
KJ thanks to the financial supported by the Hunan
Science Foundation of China (Grant No. 2015JJ3127).
We would like to thank Prof. An-Chang Shi in McMaster University for useful discussions.
We also would like to thank the referees for their thoughtful
comments which are helpful for us to improve our article.


\end{document}